\documentclass[aps,superscriptaddress,reprint,showpacs,floatfix]{revtex4-2}

\usepackage{graphicx}
\usepackage{amsmath}
\usepackage{xcolor}

\newcommand{\ghost}[1]{}

\newcommand{\tadd}[1]{{\color{black}#1}}

\newcommand{\abinitio}{\textit{ab initio}}
\newcommand{\etal}{\textit{et al.}}

\newcommand{\tcs}{ThCr$_2$Si$_2$}
\newcommand{\tc}{$T_\mathrm{c}$}

\newcommand{\tccalc}{$T_\mathrm{c}^\mathrm{calc}$}
\newcommand{\tcexpt}{$T_\mathrm{c}^\mathrm{expt}$}
\newcommand{\omegaln}{{$\omega_{\ln}$}}

\newcommand{\NumDatabase}{1883}
\newcommand{\NumNonmag}{102}
\newcommand{\NumUnstable}{36}
\newcommand{\NumStable}{66}
\newcommand{\NumReported}{24}
\newcommand{\NumUnreported}{42}

\begin{document}
\title{\textit{Ab initio} screening for BCS-type superconductivity in {\tcs}-type compounds}
\author{Tom Ichibha}
\affiliation{
  School of Information Science, JAIST,
  Asahidai 1-1, Nomi, Ishikawa 923-1292, Japan
}
\email{ichibha@gmail.com}
\author{Ryo Maezono}
\affiliation{
  \tadd{
  Graduate Major in Materials and Information Sciences,
  Institute of Science Tokyo,
  2-12-1-S6-22 Ookayama, Meguro-ku,
  Tokyo 152-8550, Japan.
  }
}
\author{Kenta Hongo}
\affiliation{
  Research Center for Advanced Computing Infrastructure,
  JAIST, Asahidai 1-1, Nomi, Ishikawa 923-1292, Japan
}
\begin{abstract}
  In this study, we applied {\abinitio} {\tc} calculations to compounds
  with the {\tcs}-type structure to search for BCS superconductor candidates.
  From the {\NumDatabase} compounds registered
  in the Inorganic Crystal Structure Database,
  we excluded those whose chemical compositions would inhibit the emergence
  of BCS-type superconductivity by giving rise to magnetism
  or heavy-fermionic behavior.
  We then focused on {\NumStable} compounds confirmed
  to be dynamically stable through phonon calculations.
  Among these, for the {\NumReported} systems
  with experimentally reported {\tc} values,
  we verified that the {\abinitio} {\tc} calculations
  exhibit excellent predictive reliability.
  For the remaining {\NumUnreported} compounds lacking experimental {\tc} values,
  our predictions identified several new BCS-type superconductor candidates,
  including SrPb$_2$Al$_2$ $\left(T_c^\mathrm{calc}=2.2\,\mathrm{K}\right)$.
\end{abstract}
\maketitle
\section{Introduction}
\label{sec:intro}\ghost{sec:intro}
The discovery of new superconductors remains one of the most exciting challenges
in materials science and condensed matter physics.
While exotic mechanisms of superconductivity have long attracted attention
in the quest for high superconducting transition temperatures (\tc)
~\tadd{\cite{2021Xingjiang_Eremets,2024Hayden_Tranquada,2022Alexander_Hemley}},
conventional BCS-type superconductors have recently garnered renewed interest,
particularly following the discovery of hydride superconductors~
\tadd{\cite{2024Boebinger_Vojta,2022Nekrasov_Sergei,2023Sun_Yanming,2023Zhao_Cui}}.
Accordingly, {\abinitio} calculations of {\tc} based on density functional theory (DFT)
and the Allen-Dynes formula~\cite{1975ALL} have become increasingly important
for the discovery of new superconductors.
It is important to note, however, that several classes of superconductors,
such as cuprates and iron-based superconductors, cannot be explained within the BCS framework.
Therefore, in high-throughput studies,
it is crucial to carefully exclude such materials from the candidate pool.
Taking this into account, high-throughput screening provides a powerful strategy
for discovering novel BCS-type superconductors.

\begin{figure}[htbp]
  \includegraphics[width=5.6cm]{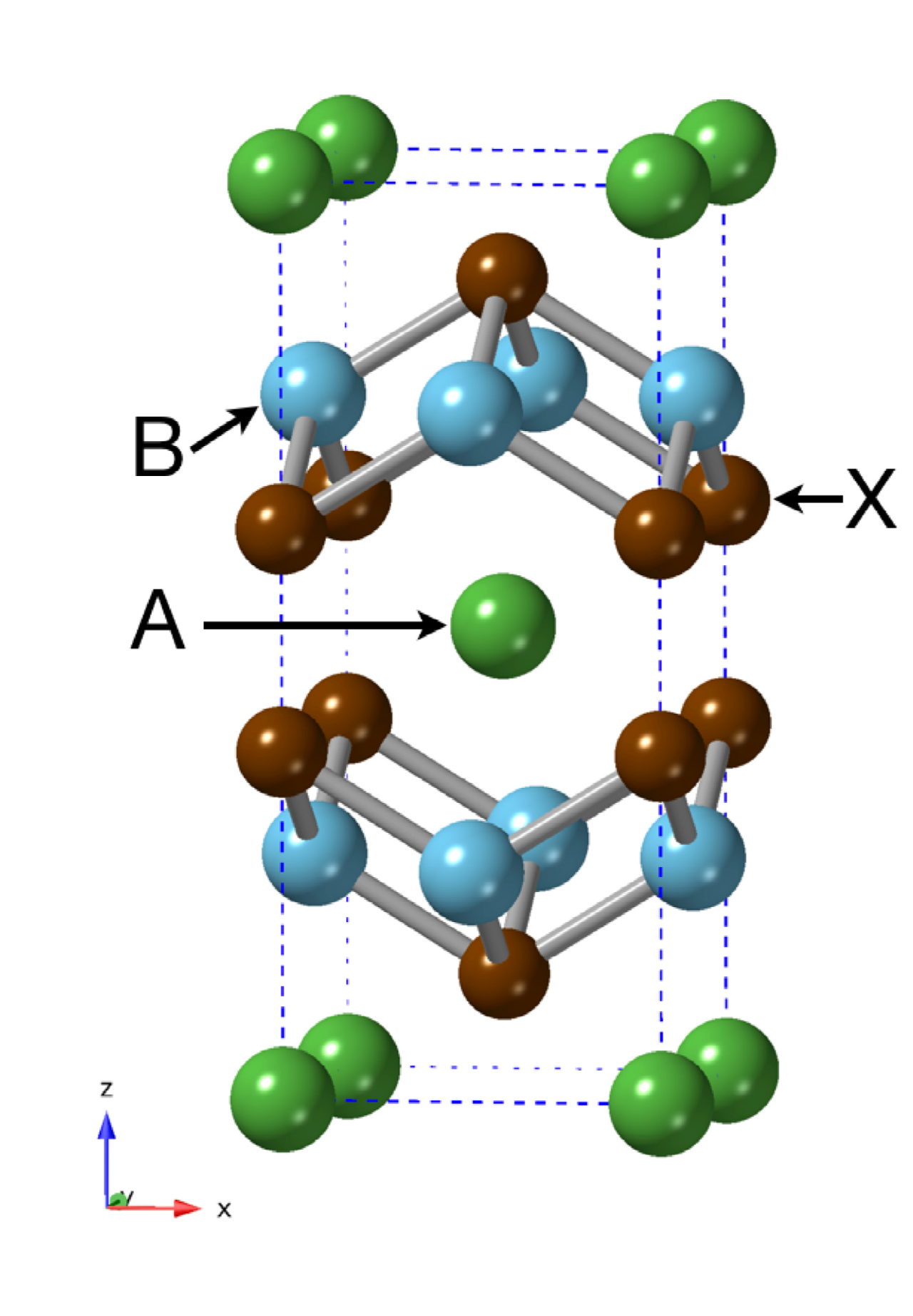}
  \caption{
    \label{fig:structure}\ghost{fig:structure}
    Conventional unit cell of $AB_{2}X_{2}$ with a tetragonal, {\tcs}-type structure.
    The space group is $I$4/$mmm$ (No. 139).
    $A$, $B$, and $X$ are a rare-earth/alkaline earth element;
    a transition metal; and an element belonging to group 15, 14,
    and occasionally 13 ($p$-block elements), respectively.
  }
\end{figure}

\vspace{2mm}
Among the most intriguing targets in the search for BCS-type superconductors
is the family of {\tcs}-type structures with space group $I4/mmm$.
This class of compounds includes both exotic superconductors,
such as high-temperature iron pnictides~\tadd{\cite{2022Fernandes_Kotliar,2016Si_Abrahams}},
and BCS-type superconductors, including iron-free pnictides\tadd{\cite{2017Zhang_Zhai,2019Shatruk}}.
Despite significant interest in these materials, many compounds in this family remain unexplored
in terms of their physical properties.
Structurally, they are characterized by a stacked-layer configuration in the order of $A$-[$X$-$B_2$-$X$]-$A$,
as illustrated in Fig.~\ref{fig:structure},
where $A$ denotes a rare-earth or alkaline-earth metal, $B$ is a transition metal, and $X$ is a group 15, 14,
or occasionally 13 $p$-block element, forming a quasi-two-dimensional (2D) network~\cite{1965BAN}.
With over 1000 possible structural variants arising from the vast combination space of $A$-$B$-$X$,
this system represents a highly promising platform for high-throughput computational studies.

\vspace{2mm}
\tadd{
  Although one possible approach is to modify the composition
  based on previously reported BCS-type superconductors with high critical temperatures {\tc},
  such as NaC$_6$\cite{2016LU_Hemley,2022Khan_Naqib}, KPbB$_6$C$_6$\cite{2023Nisha_Zurek},
  and MgIr$_2$H$_6$\cite{2024Dolui_Pickard},
  we have instead chosen to focus on the fact that {\tc} remains unreported for many {\tcs}-type compounds.
  From a fundamental scientific perspective, we aim to evaluate their superconducting potential
  by systematically exploring this largely unexplored space.
}

\vspace{2mm}
In this study, we screened new BCS superconductor candidates using {\abinitio} {\tc}
calculations based on the Allen--Dynes framework for {\NumDatabase} {\tcs}-type compounds
registered in the Inorganic Crystal Structure Database (ICSD) \cite{1983BER}.
\tadd{These compounds are synthesized experimentally.}
During the screening, compounds containing elements that possess characteristics inhibiting
the emergence of BCS superconductivity were excluded in advance,
and compounds predicted to be dynamically unstable from phonon calculations were also eliminated
(Section~\ref{ssec:screen}).
For the remaining {\NumStable} compounds, {\tc} was evaluated via {\abinitio} calculations.
Prior to exploring new candidates, validation on {\NumReported} compounds
with reported experimental {\tc} values demonstrated that the {\abinitio} {\tc} calculations
reliably assess the BCS superconducting transition temperatures of {\tcs}-type compounds
(Section~\ref{ssec:validate}).
Consequently, by applying the {\abinitio} {\tc} calculations to the remaining {\NumUnreported} compounds
lacking experimental {\tc} reports, we discovered several new candidate BCS superconductors,
including SrPb$_2$Al$_2$ ({\tccalc}=2.2~K) (Section~\ref{ssec:unreported}).

\section{Theory}
\label{sec:theory}\ghost{sec:theory}
The $T_{c}$ values were evaluated using the Allen-Dynes formula \cite{1975ALL}:
\begin{equation}
  T_{c}=\cfrac{\omega_{\rm ln}}{1.2}
  \exp \left[-\cfrac{1.04(1+\lambda)} {\lambda-\mu^{*}(1+0.62 \lambda)}\right].
  \label{eq:allen-dynes}
\end{equation}\ghost{eq:allen-dynes}
Here, {\omegaln} is the logarithmic average phonon frequency,
$\lambda$ is the electron-phonon coupling constant, and $\mu^{*}$ is the effective Coulomb repulsion.
In this study, we adopted $\mu^{*} = 0.1$, which is considered a reasonable value
for nearly-free-electron metals\cite{1968MCM,1975ALL}.
The quantities $\lambda$ and {\omegaln} are given by the following equations:
\begin{align}
\lambda &=2 \int_0^{\infty}d\omega\cdot\frac{\alpha\left(\omega\right)^2F\left(\omega\right)}{\omega},
\label{eq.lambda} \\
\omega_{\ln} &= \exp \left( \frac{2}{\lambda} \int_0^\infty d\omega \frac{\alpha(\omega)^2F(\omega) \ln \omega}{\omega} \right).\label{eq:omega-ln}
\end{align}
\ghost{eq:lambda,eq:omega-ln}
The Eliashberg function $\alpha(\omega)^2F(\omega)$ represents the total strength
of the coupling between electrons on the Fermi surface and phonons with frequency $\omega$:
\begin{align}
  \alpha\left(\omega\right)^2F\left(\omega\right) &= \frac{1}{N\left(\varepsilon_{F}\right)}\sum_{\mathbf{q}\nu}
  \delta\left(\omega-\omega_{\mathbf{q}\nu}\right)
  \sum_{\mathbf{k}mn}\left|g_{mn,\nu}\left(\mathbf{k},\mathbf{q}\right)\right|^2
  \nonumber \\ &\cdot
  \delta\left(\varepsilon_{\mathbf{k}+\mathbf{q},\,m}-\varepsilon_{F}\right)\delta\left(\varepsilon_{\mathbf{k},\,n}-\varepsilon_{F}\right),
  \label{eq:eliaschberg}\\
  g_{mn,\nu}\left(\mathbf{k},\mathbf{q}\right) &= \frac{1}{\sqrt{2\omega_{\mathbf{q}\nu}}}\left\langle\psi_{\mathbf{k}+\mathbf{q},\,m}\left(\mathbf{r}\right)\left|\frac{\partial v_{\mathrm{eff}}\left(\mathbf{r}\right)}{\partial\mathbf{r}_{\mathbf{q},\,\nu}}\right|\psi_{\mathbf{k},\,n}\left(\mathbf{r}\right)\right\rangle.
  \label{eq:el-ph-matrix}
\end{align}
\ghost{eq:eliaschberg, eq:el-ph-matrix}
Here, $\varepsilon_{F}$ is the Fermi energy and $N\left(\varepsilon_{F}\right)$ is the density of states at the Fermi level.
$v_{\mathrm{eff}}(\mathbf{r})$ is the effective potential acting on the Kohn-Sham orbitals,
and $\partial v_{\mathrm{eff}}(\mathbf{r}) / \partial \mathbf{r}_{\mathbf{q},\,\nu}$ is its derivative
along the displacement direction of the vibrational mode $\mathbf{q},\,\nu$.

\section{Calculation details}
\label{sec:details}\ghost{sec:details}
\tadd{
  We screened new BCS conductors from {\NumDatabase} {\tcs}-type compounds listed
  in the ICSD database \cite{1983BER} following the workflow described in section \ref{ssec:screen}.
}
After structural optimization, the dynamic stability of the system was assessed
based on harmonic phonon calculations.
For the dynamically stable systems, {\tc} was evaluated according to the theory described
in the previous section \ref{sec:theory}.

\vspace{2mm}
We used density functional theory implemented in the Quantum Espresso (QE) package \cite{2009GIA}
was used with the PBE exchange--correlation functional \cite{1999GKR,1994GKR,1996GKR}.
The core electrons were replaced by the Vanderbilt-type ultrasoft pseudopotentials \cite{1990DV}
provided in the PS library \cite{2014DAL}.
The plane wave energy cut-off was 100 Ry, and $6 \times 6 \times 6$ $k$-mesh was used for every calculation.
Since the calculated {\tcs}-type compounds were metallic,
the Marzari-Vanderbilt cold smearing scheme~\cite{1999MAR}
with a broadening width of 0.02 Ry was applied to all the compounds.
Self-consistent field (SCF) iterations were continued until the total energy difference
between iterations getting below $1.0 \times 10^{-8}$~Ry.
For the geometry optimization, both lattice parameters and atomic positions were relaxed
until the total energy difference between ionic steps getting below $1.0 \times 10^{-4}$ Ha
and every ionic force getting below $1.0 \times 10^{-3}$~$\mathrm{Ha}/a_0$.
For the phonon properties, the density functional perturbation theory (DFPT) was used
with a $6 \times 6 \times 6$ $q$-mesh same as the $k$-mesh.
\tadd{
  Based on the wave functions obtained by QE, crystal orbital Hamilton population (COHP) analyses
  \cite{1993Dronskowski_Bloechl, 2011Deringer_Dronskowski} were conducted using the LOBSTER code
  \cite{2013Maintz_Dronskowski,2016Maintz_Dronskowski,2020Nelson_Dronskowski,2016Maintz_Dronskowski}.
  \tadd{Fermi surfaces were visualized using FermiSurfer\cite{2019Kawamura}.}
}

\vspace{2mm}
For the {\tc} calculation, we employed a double-grid technique~\cite{2006WIERZBOWSKA_GIANNOZZI}
to evaluate the $k$- and $q$-space integrals in Eq.~\eqref{eq:eliaschberg}.
The orbital energies $\varepsilon_{\mathbf{k},m}$ were calculated using
a dense $36 \times 36 \times 36$ $k$-mesh.
Phonon frequencies $\omega_{\mathbf{q}\nu}$ and electron-phonon matrix elements
$\left| g_{mn,\nu}(\mathbf{k}, \mathbf{q}) \right|$, which are originally obtained
on a coarser $q$-grid via DFPT,
were interpolated onto the same dense grid using Fourier interpolation.

\section{Results and discussion}
\label{sec:results}\ghost{sec:results}
\subsection{Screening dynamically stable compounds}
\label{ssec:screen}\ghost{ssec:screen}
The {\tcs}-type structures of {\NumDatabase} compounds are available
in the ICSD database \cite{1983BER}.
\tadd{The compounds are synthesized experimentally.}
Some of these compounds were known to potentially display magnetic orderings
(i.e., Cr, Mn, Fe, Co, and Ni-based compounds) and heavy-fermionic behavior
(i.e., displaying an $f$ electron in their valence state),
which prevent their use as conventional superconductors.
Thus, these compounds were pre-filtered out in this study,
leading to further reduction of the pool to {\NumNonmag} compounds.

\vspace{2mm}
Furthermore, for the remaining compounds, we performed phonon calculations
to identify dynamically stable structures and then evaluated {\tc} for those as well.
\tadd{
  In general, when delving into a specific system,
  it is possible to consider not only dynamic stability but also kinetic stability,
  as assessed by molecular dynamics simulations, and the possibility of reaction pathways.
  However, since the main focus of this study is not the analysis of individual materials
  but rather high-throughput candidate screening across a broad range of systems,
  such analyses were not performed here in order to maintain high throughput.
}
The phonon calculations revealed that {\NumStable} compounds were dynamically stable,
while other {\NumUnstable} compounds were not.
Table~\ref{tab.NegMode} lists up some of the compounds that exhibits imaginary frequencies.

\vspace{2mm}
The presence of unstable structures is interesting
because these compounds have been experimentally synthesized,
indicating that they should be dynamically and energetically stable under ambient conditions.
The discrepancy can be interpreted as follows.
For Pt-based compounds, several studies have proposed that a very careful
study was needed for identifying the symmetry of the crystal structure,
i.e., either $I$4/$mmm$ (No. 139) or $P$4/$nmm$ (No. 129).
For example, Venturini {\etal} determined the crystal
structure of LaPt$_2$Ge$_2$ using X-ray diffraction~\cite{1989VEN}.
They found that the monoclinic LaPt$_2$Ge$_2$ structure
was closer to the CaBe$_2$Ge$_2$-type structure ($P$4/$nmm$),
rather than the ThCr$_2$Si$_2$-type structure ($I$4/$mmm$).
For ThPt$_2$Ge$_2$, two different space groups have been reported.
In 1977, Marazza {\etal} \cite{1977MAR} claimed that
ThPt$_2$Ge$_2$ possessed the $I$4/$mmm$ space group by X-ray diffraction.
On the contrary, in 1984, Shelton {\etal} proposed
that the crystal structure was $P$4/$nmm$~\cite{1984SHE}.
Similarly, for LaPt$_2$Si$_2$,
Hase {\etal} found that the CaBe$_2$Ge$_2$ structure was more stable than
the ThCr$_2$Si$_2$-type ($I$4/$mmm$) structure by $\sim$25~mRy~\cite{2013HAS} from a DFT study,
implying that the CaBe$_2$Ge$_2$-type ($P$4/$nmm$) structure is the true space group.
On the other hand, CaCu$_2$P$_2$ actually possesses $I$4/$mmm$,
unlike the above Pt-based compounds, while it has been recently
proposed that a ferromagnetic phase is potentially
more stable than the paramagnetic phase ~\cite{2020ZAD}.
Thus, the literature studies summarized in Table~\ref{tab:unstable} implied
that the 4 compounds with $I$4/$mmm$ were dynamically unstable because
either a different crystal structure (i.e., $P$4/$nmm$)
was the true space group, or a magnetic ordering phase
(i.e., ferromagnetic) was the true ground state.

\vspace{2mm}
\tadd{
  When imaginary phonon modes are observed, further analyses can be conducted
  if there is a particular interest in a specific material system.
  For example, one could identify a lower-symmetry phase distorted
  along the direction of the unstable mode and compare it
  with experimental XRD peaks~\cite{2016Nakano_Maezono,2017Nakano_Maezono}.
  However, since the main focus of this study is high-throughput candidate screening,
  we have chosen to exclude materials exhibiting imaginary modes
  from the candidate pool and do not pursue such analyses here.
  Detailed discussions are provided for four typical examples,
  while the remaining 32 cases are similarly regarded as likely distorted
  to lower symmetry and thus excluded from the high-throughput screening.
}

\begin{table*}[htb]
  \caption{
    \label{tab:unstable}\ghost{tab:unstable}
    Four compounds that exhibited imaginary frequency modes. SG denotes the space group.
  }
    \renewcommand{\arraystretch}{1.3}
    \begin{tabular}{l p{6cm} l l }
      \hline
      Compound & Negative region & Possible reason & References{\footnotemark[1]}\\
      \hline
      LaPt$_2$Ge$_2$   & Z to N, between N--P, X--$\Gamma$ & $P$4/$nmm$ is the true SG. & Ref.~\onlinecite{1977MAR}, \onlinecite{1984SHE} \\
      ThPt$_2$Ge$_2$   & Z to N, between N--P, X--$\Gamma$ & $P$4/$nmm$ is the true SG. & Ref.~\onlinecite{1989VEN} \\
      LaPt$_2$Si$_2$   & between $\Gamma$--N, X--$\Gamma$  & $P$4/$nmm$ is the true SG. & Ref.~\onlinecite{2013HAS} \\
      CaCu$_2$P$_2$    & Z, $\Gamma$ to N & Ferromagnetic & Ref.~\onlinecite{2020ZAD} \\
      \hline
    \end{tabular}
    \footnotetext[1]{
      References that propose a crystal structure different from $I$4/$mmm$
      or a possible magnetic ordering (i.e, ferromagnetic). See the main text.
    }
  \label{tab.NegMode}
\end{table*}

\subsection{Validation of {\abinitio} {\tc} evaluation}
\label{ssec:validate}
\begin{figure}[htbp]
    \includegraphics[width=\hsize]{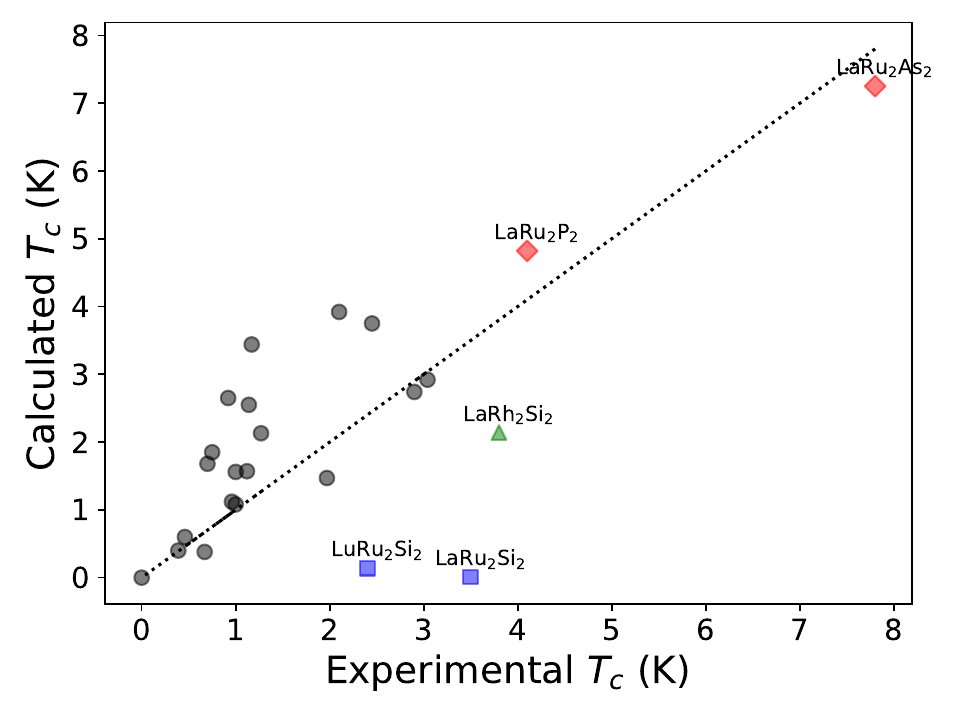}
  \caption{
  \label{fig:compare-tc}\ghost{fig:compare-tc}
    Comparison of calculated and experimental $T_{c}$ for \tadd{{\NumReported}} compounds.
    For LaRh$_2$Si$_2$, indicated by a green triangle, two different experimental {\tc} values—3.8~K \cite{1983IF_IN}
    and 0.074~K \cite{1986TTMP_JAM}—have been reported, as shown in Table~\ref{tab:reported}.
    Our calculated result is more consistent with the 3.8~K value, and this value is used in Fig.~\ref{fig:compare-tc}.
  }
\end{figure}

\ghost{ssec:validate}
Within the dynamically stable {\NumStable} compounds,
{\NumReported} compounds have corresponding experimental {\tc},
while the {\tc} values of the rest {\NumUnreported} compounds have not been reported.
To find novel superconductors among these {\NumUnreported} compounds,
the {\abinitio} {\tc} calculations were initially validated
using the {\NumReported} known compounds.
Fig.~\ref{fig:compare-tc} shows the comparison between our estimations of {\tc} and experimental data.
The numerical data is given in Table~\ref{tab:reported}.
The calculated {\tc} values were in reasonable agreement with the experimental values,
with the Pearson correlation coefficient equaling 0.71.
In particular, it is remarkable that our calculations successfully reproduced the {\tc}
of LaRu$_2$As$_2$ ({\tcexpt} = 7.8K)\cite{2016QG_ZAR}
\footnote{
  LaRu$_2$As$_2$ has been suggested to be a multi-band superconductor
  with multiple bands contributing to superconductivity,
  as indicated by DFT calculations \cite{2017Hadi_Islam}.
}
and LaRu$_2$P$_2$ ({\tcexpt} = 4.1K)~\cite{1987WJ_LB},
which are shown as red diamonds and exhibit the highest experimental {\tc} values.

\vspace{2mm}
On the other hand, it is an intriguing discrepancy that LaRu$_2$Si$_2$ and LuRu$_2$Si$_2$,
indicated by blue squares, exhibit nearly 0~K in the calculated {\tc} values,
despite relatively high experimental {\tc} having been reported.
This suggests that superconductivity in these systems is not driven by the BCS mechanism.
Indeed, enhanced itinerant electron paramagnetism has been observed in these compounds \cite{1984IF_IN},
which indicates that they are close to the critical condition for spontaneous magnetization—the Stoner criterion.
Therefore, it is possible that in these systems, Cooper pairs are formed not
via electron-phonon coupling but rather through spin fluctuations.

\vspace{2mm}
When excluding these two compounds, LaRu$_2$Si$_2$ and LuRu$_2$Si$_2$,
which are likely non-BCS-type superconductors, the resulting correlation coefficient is 0.86,
indicating the sufficient reliability of our {\tc} evaluations for BCS-type superconductors.
\tadd{
  Except for the two compounds and LaRh$_2$Si$_2$ by the green triangle,
  the calculated {\tc} provides an upper bound for the experimental {\tc}.
  One possible reason is that, experimentally, the observed {\tc} may be
  lower than the ideal limit due to the influence of impurities
  and point, line, or planar defects.
  Another reason is related to the effective Coulomb repulsion $\mu^{*}$
  in Eq.~\eqref{eq:allen-dynes}; as described in Section~\ref{sec:theory},
  we used $\mu^{*}=0.1$, which is the lower end of the typical range.
  In LaRh$_2$Si$_2$, a long-range magnetic order of itinerant-electron origin
  coexists with the superconducting state \cite{1983IF_IN}.
  Such an anomaly could be a reason for the underestimation.
}

\begin{table*}[htbp]
  \caption{
    \label{tab:reported}\ghost{tab:reported}
    Comparison between the calculated and experimental {\tc} values for the {\NumReported} compounds with reported experimental {\tc}.
    Key physical quantities required for the {\tc} calculations are also listed.
  }
  \begin{tabular}{lccccc|lccccc}
    \hline
    Compound & $N_\mathrm{F}$ \tadd{(states/eV)} & {\omegaln} (K) & $\lambda$ & {\tccalc} (K) & {\tcexpt} (K) &
    Compound & $N_\mathrm{F}$ \tadd{(states/eV)} & {\omegaln} (K) & $\lambda$ & {\tccalc} (K) & {\tcexpt} (K) \\
    \hline
    LaRu$_2$As$_2$ & 2.2 & 90.22  & 1.10 & 7.25 & 7.8\cite{2016QG_ZAR} &
    YPd$_2$Ge$_2$	 & 2.6 & 114.94 & 0.57 & 2.55 & 1.14\cite{2018CHA} \\
    LaRu$_2$P$_2$	 & 2.1 & 143.50 & 0.69 & 4.82 & 4.1\cite{1987WJ_LB} &
    LaPd$_2$Ge$_2$ & 2.9 & 104.76 & 0.63 & 1.57 & 1.12\cite{1981GWH_JEB} \\
    LaRh$_2$Si$_2$ & 4.0 & 252.71 & 0.46 & 2.14 & 3.8\cite{1983IF_IN}, 0.074\cite{1986TTMP_JAM} &
    BaRh$_2$P$_2$	 & 4.5 & 188.18 & 0.55 & 1.56 & 1.0\cite{2010DH_HT} \\
    LaRu$_2$Si$_2$ & 2.5 & 224.40 & 0.25 & 0.01 & 3.5\cite{1984IF_IN} &
    CaPd$_2$P$_2$	 & 8.4 & 165.06 & 0.43 & 1.08 & 1.0\cite{2020JB_WX} \\
    SrPd$_2$Ge$_2$ & 2.4 & 89.80  & 0.80 & 2.92 & 3.04\cite{2009HF_AS} &
    LaPd$_2$P$_2$	 & 1.0 & 165.09 & 0.44 & 1.12 & 0.96\cite{2016GD_PCC} \\
    SrIr$_2$As$_2$ & 3.6 & 87.90  & 0.67 & 2.74 & 2.9\cite{2010DH_HT} &
    SrPd$_2$As$_2$ & 2.6 & 106.66 & 0.56 & 2.65 & 0.92\cite{2013VKA_DCJ} \\
    BaIr$_2$As$_2$ & 4.5 & 121.93 & 0.67 & 3.75 & 2.45 \cite{2017WANG_REN} &
    YPd$_2$P$_2$	 & 0.9 & 122.97 & 0.52 & 1.85 & 0.75\cite{2016GD_PCC} \\
    LuRu$_2$Si$_2$ & 3.2 & 186.72 & 0.31 & 0.14 & 2.4\cite{1984IF_IN} &
    SrPd$_2$P$_2$	 & 3.0 & 173.04 & 0.47 & 1.68 & 0.7\cite{2020JB_WX} \\
    BaIr$_2$P$_2$	 & 4.0 & 189.77 & 0.58 & 3.92 & 2.1\cite{2010DH_HT} &
    LuPd$_2$Si$_2$ & 2.3 & 161.08 & 0.45 & 0.38 & 0.67\cite{1986TTMP_JAM} \\
    CaPd$_2$Ge$_2$ & 2.0 & 104.33 & 0.62 & 1.47 & 1.97 \cite{2014ANAND_JOHNSTON} &
    YPd$_2$Si$_2$	 & 1.8 & 169.48 & 0.39 & 0.6  & 0.46\cite{2019GC_DK} \\
    CaPd$_2$As$_2$ & 2.5 & 96.95  & 0.57 & 2.13 & 1.27\cite{2013VKA_DCJ} &
    LaPd$_2$Si$_2$ & 2.2 & 166.54 & 0.36 & 0.4  & 0.39\cite{1986TTMP_JAM} \\
    YbPd$_2$Ge$_2$ & 2.1 & 93.69  & 0.72 & 3.44 & 1.17\cite{1981GWH_JEB} &
    LiCu$_2$P$_2$	 & 0.7 & 199.93 & 0.23 & 0.0  & 0.0\cite{2011FH_HHW} \\
    \hline
  \end{tabular}
\end{table*}
\begin{table*}[htbp]
  \caption{
    \label{tab:unreported}\ghost{tab:unreported}
    Prediction of {\tc} for the {\NumUnreported} compounds for which experimental {\tc} values have not been reported,
    along with key physical quantities required for the {\tc} calculations.
  }
    \begin{tabular}{lcccc|lcccc}
      \hline
      Compound & $N_\mathrm{F}$ \tadd{(states/eV)} & \omegaln (K) & $\lambda$ & {\tccalc} (K) &
      Compound & $N_\mathrm{F}$ \tadd{(states/eV)} & \omegaln (K) & $\lambda$ & {\tccalc} (K) \\
      \hline
      SrPb$_2$Al$_2$ & 2.4 & 62.97  & 0.69 & 2.152 &
      ThRu$_2$Si$_2$ & 2.5 & 175.15 & 0.32 & 0.166 \\
      CaPd$_2$Si$_2$ & 2.6 & 201.32 & 0.46 & 1.781 &
      YRh$_2$Si$_2$  & 3.4 & 266.01 & 0.28 & 0.156 \\
      BaCu$_2$In$_2$ & 3.4 & 95.63  & 0.54 & 1.580 &
      CaCu$_2$Ge$_2$ & 1.6 & 169.15 & 0.32 & 0.142 \\
      SrOs$_2$P$_2$  & 1.9 & 118.73 & 0.50 & 1.402 &
      CaAg$_2$Ge$_2$ & 1.6 & 141.16 & 0.31 & 0.101 \\
      YRh$_2$Ge$_2$  & 2.7 & 156.56 & 0.45 & 1.197 &
      BaRu$_2$Sb$_2$ & 1.9 & 138.97 & 0.30 & 0.074 \\
      SrAu$_2$Si$_2$ & 1.6 & 122.83 & 0.45 & 0.981 &
      LaOs$_2$Si$_2$ & 2.5 & 184.10 & 0.29 & 0.062 \\
      ThOs$_2$Ge$_2$ & 3.0 & 131.74 & 0.44 & 0.942 &
      YRu$_2$Si$_2$  & 2.9 & 235.52 & 0.28 & 0.054 \\
      ThAu$_2$Si$_2$ & 1.8 & 131.88 & 0.42 & 0.707 &
      BaOs$_2$P$_2$  & 2.0 & 145.82 & 0.29 & 0.042 \\
      SrCu$_2$Ge$_2$ & 1.6 & 141.00 & 0.40 & 0.616 &
      BaZn$_2$Si$_2$ & 0.8 & 199.21 & 0.28 & 0.041 \\
      LaPt$_2$Si$_2$ & 2.2 & 129.50 & 0.41 & 0.613 &
      CaCu$_2$Si$_2$ & 1.6 & 238.13 & 0.27 & 0.029 \\
      CaAu$_2$Si$_2$ & 1.5 & 142.04 & 0.40 & 0.601 &
      LaZn$_2$Al$_2$ & 1.7 & 187.77 & 0.27 & 0.026 \\
      BaAg$_2$Ge$_2$ & 1.8 & 123.15 & 0.40 & 0.521 &
      ThCu$_2$Si$_2$ & 1.5 & 177.76 & 0.27 & 0.022 \\
      LaRh$_2$Ge$_2$ & 2.9 & 182.91 & 0.37 & 0.472 &
      LaCu$_2$Si$_2$ & 1.5 & 248.70 & 0.26 & 0.020 \\
      BaZn$_2$Ge$_2$ & 1.0 & 102.19 & 0.40 & 0.424 &
      ThRh$_2$Ge$_2$ & 3.8 & 167.98 & 0.41 & 0.20  \\
      SrCu$_2$In$_2$ & 2.5 & 114.07 & 0.38 & 0.378 &
      YIr$_2$As$_2$  & 2.8 & 226.67 & 0.32 & 0.19  \\
      BaMg$_2$Ge$_2$ & 1.8 & 152.57 & 0.37 & 0.373 &
      LaAg$_2$Si$_2$ & 1.4 & 196.09 & 0.24 & 0.007 \\
      ThOs$_2$Si$_2$ & 2.6 & 165.02 & 0.36 & 0.37  &
      SrIr$_2$Ge$_2$ & 1.3 & 150.73 & 0.24 & 0.005 \\
      ThRu$_2$Ge$_2$ & 2.6 & 127.07 & 0.37 & 0.330 &
      KCu$_2$Se$_2$  & 3.0 & 106.83 & 0.24 & 0.004 \\
      ThRh$_2$Si$_2$ & 2.8 & 236.41 & 0.33 & 0.259 &
      BaRu$_2$As$_2$ & 1.0 & 153.0  & 0.22 & 0.001 \\
      SrAg$_2$Ge$_2$ & 1.6 & 127.76 & 0.34 & 0.183 &
      ScCu$_2$Si$_2$ & 1.2 & 246.13 & 0.16 & 0.000 \\
      LaCu$_2$Ge$_2$ & 1.5 & 128.68 & 0.34 & 0.171 &
      BaRh$_2$Ge$_2$ & 1.9 & 147.19 & 0.19 & 0.000 \\
      \hline
    \end{tabular}
\end{table*}
\begin{figure}[htbp]
    \includegraphics[width=\hsize]{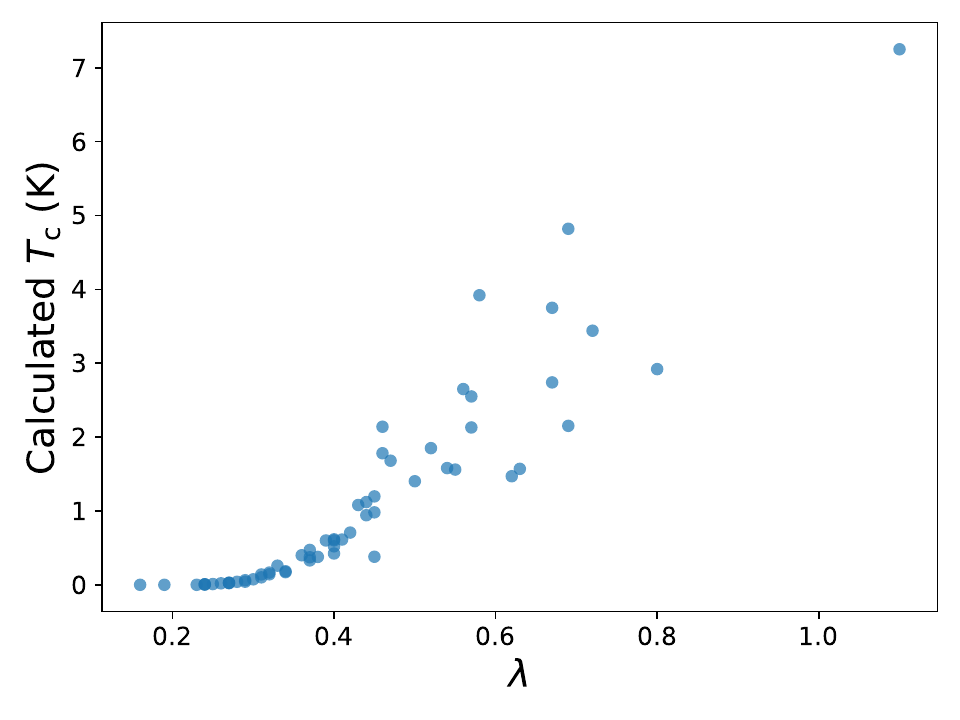}
  \caption{
  \label{fig:lambda_vs_tc}\ghost{fig:lambda\_vs\_tc}
  \tadd{
    Correlation relationship between
    electron-phonon coupling constant $\lambda$
    and calculated {\tc} of dynamically
    stable {\NumStable} structures listed
    in Tables \ref{tab:reported} and \ref{tab:unreported}.
  }
  }
\end{figure}

\subsection{\textit{Ab initio} {\tc} predictions}
\label{ssec:unreported}\ghost{ssec:unreported}
Table~\ref{tab:unreported} presents the {\abinitio} {\tc} predictions
for compounds without reported experimental {\tc} values.
The range of $\lambda$ values falls within that typically observed
for conventional metallic superconductors,
and as is well known in such cases, {\tccalc} tends to be predominantly determined by $\lambda$.
Indeed, when combining the results from Tables~\ref{tab:reported} and \ref{tab:unreported},
the Pearson correlation coefficient between $\lambda$ and {\tccalc} is found to be as high as 0.92
\tadd{(see Fig.~\ref{fig:lambda_vs_tc})}.
Among the compounds with unreported {\tcexpt}, the highest predicted {\tccalc}\tadd{=2.2~K} is
for SrPb$_2$Al$_2$,
which exhibits a $\lambda$ value of 0.69, comparable to that of
LaRu$_2$P$_2$ ({\tcexpt} = 4.1~K, {\tccalc} = 4.82~K).

\begin{figure*}[htbp]
  \begin{minipage}{0.195\hsize}
    (a)~LaRu$_2$As$_2$
  \end{minipage}
  \begin{minipage}{0.195\hsize}
    (b)~LaRu$_2$P$_2$
  \end{minipage}
  \begin{minipage}{0.195\hsize}
    (c)~SrPb$_2$Al$_2$
  \end{minipage}
  \begin{minipage}{0.195\hsize}
    (d)~CaPd$_2$Si$_2$
  \end{minipage}
  \begin{minipage}{0.195\hsize}
    (e)~CaPd$_2$P$_2$
  \end{minipage}
  \includegraphics[width=0.195\hsize]{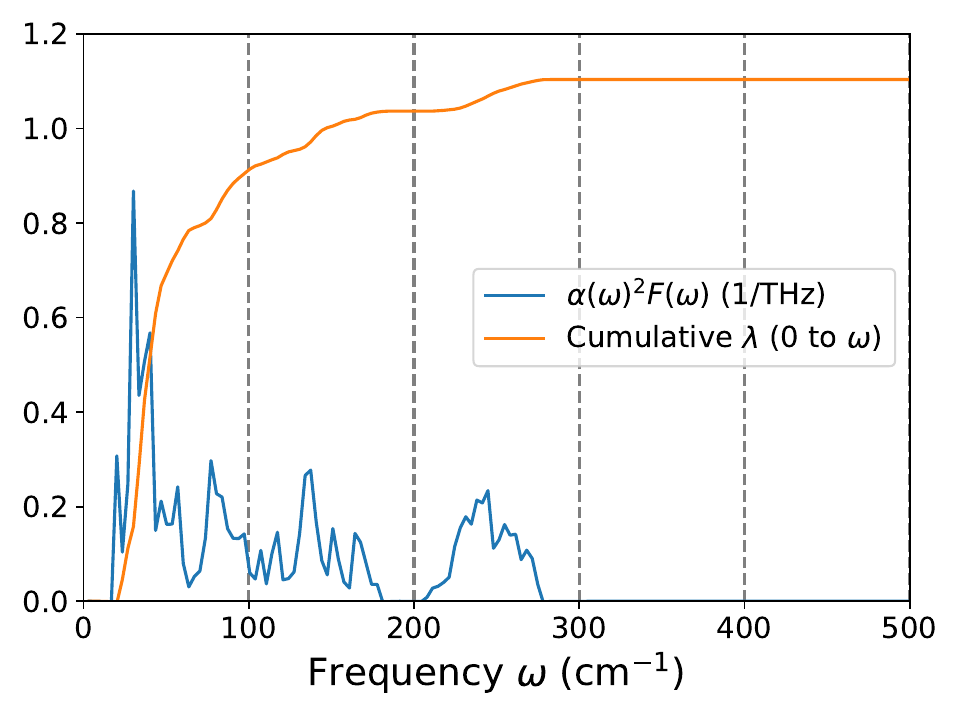}
  \includegraphics[width=0.195\hsize]{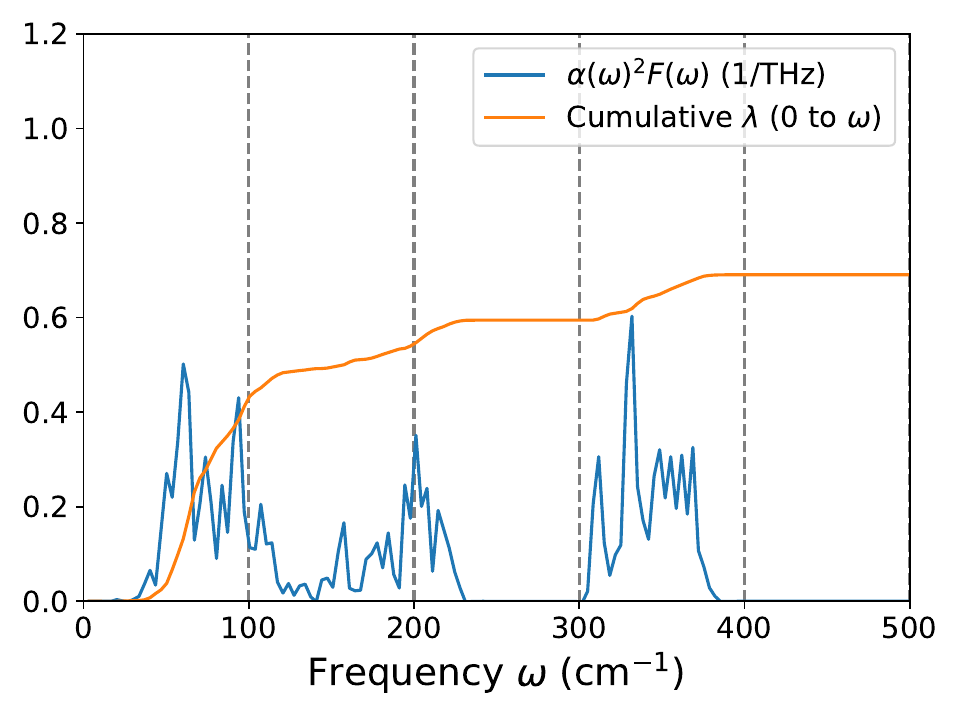}
  \includegraphics[width=0.195\hsize]{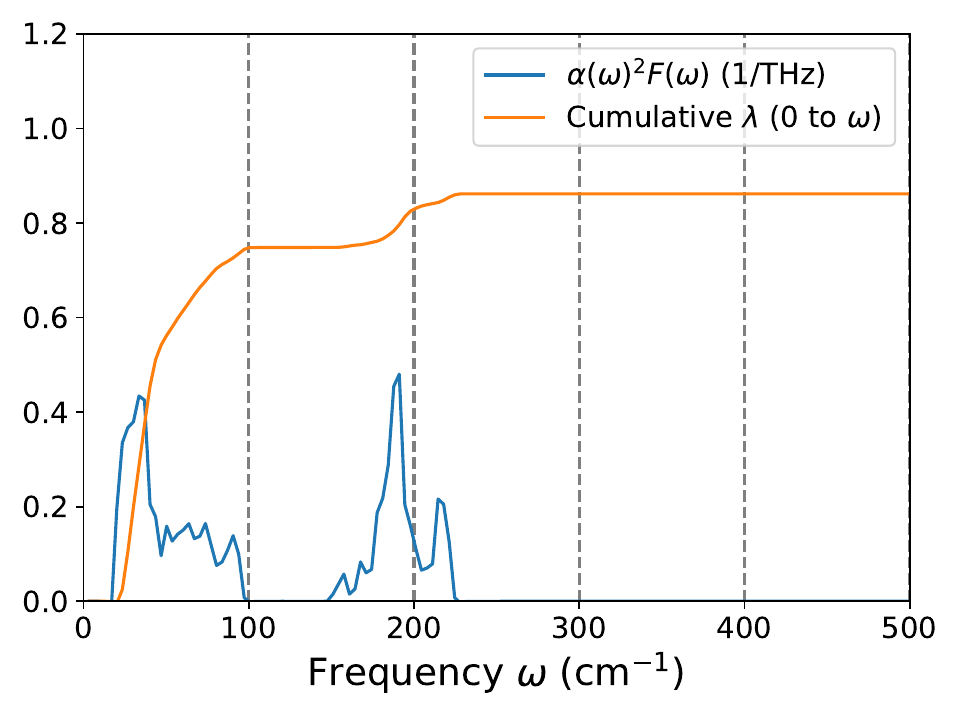}
  \includegraphics[width=0.195\hsize]{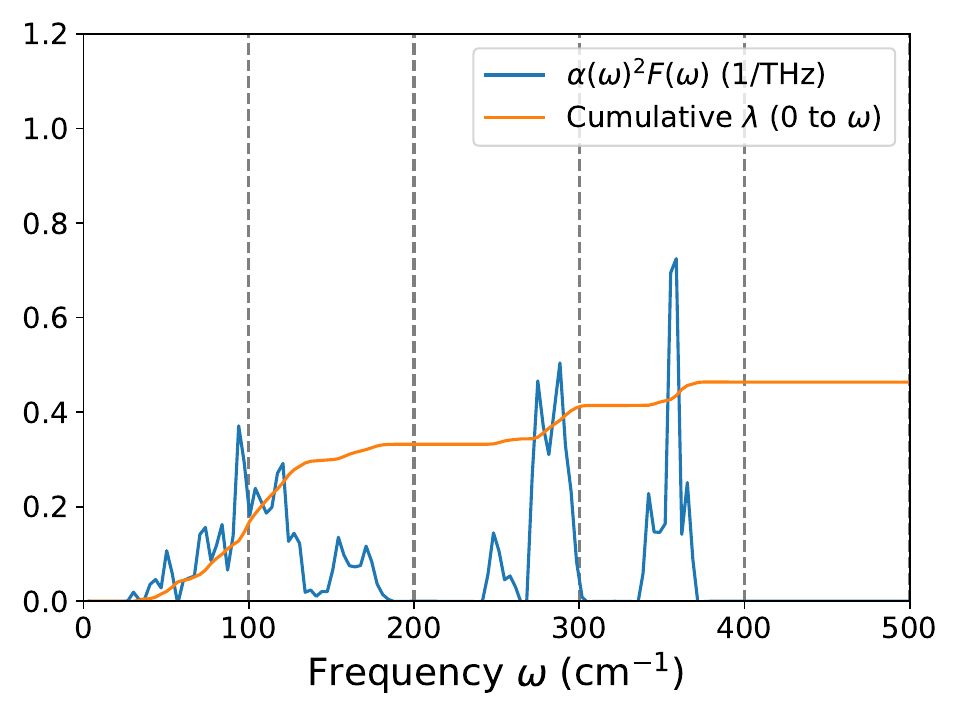}
  \includegraphics[width=0.195\hsize]{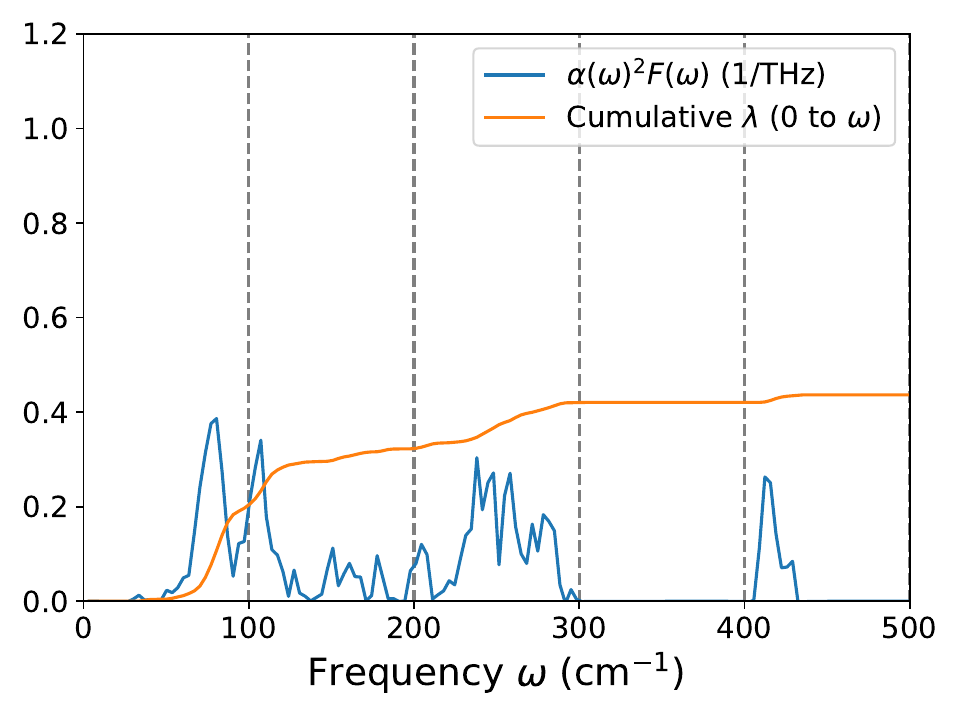}
  \\ 
  \includegraphics[width=0.195\hsize]{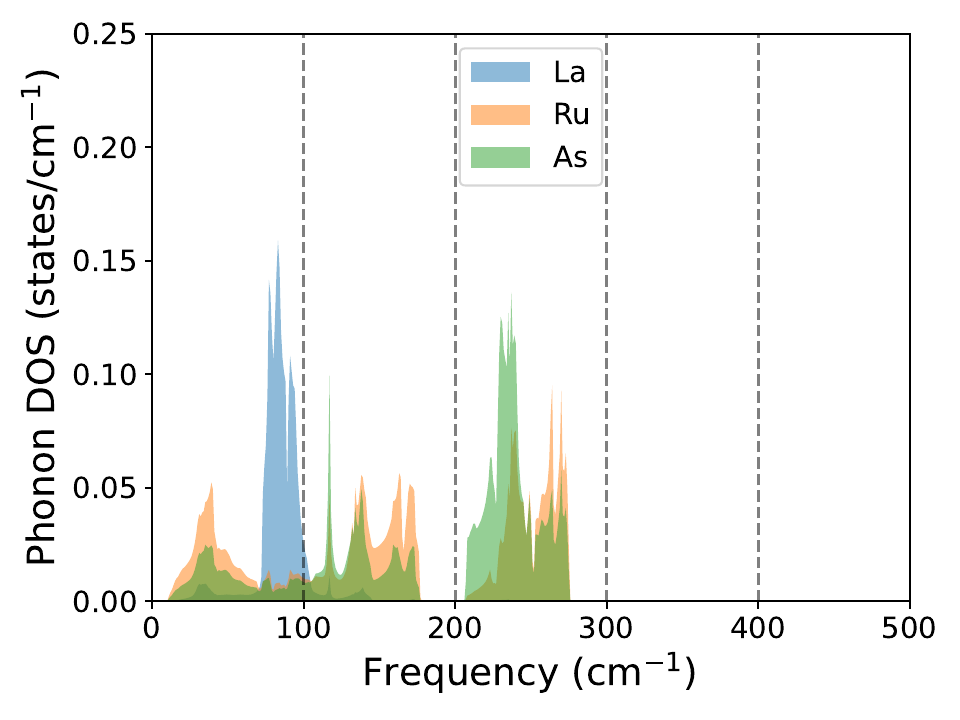}
  \includegraphics[width=0.195\hsize]{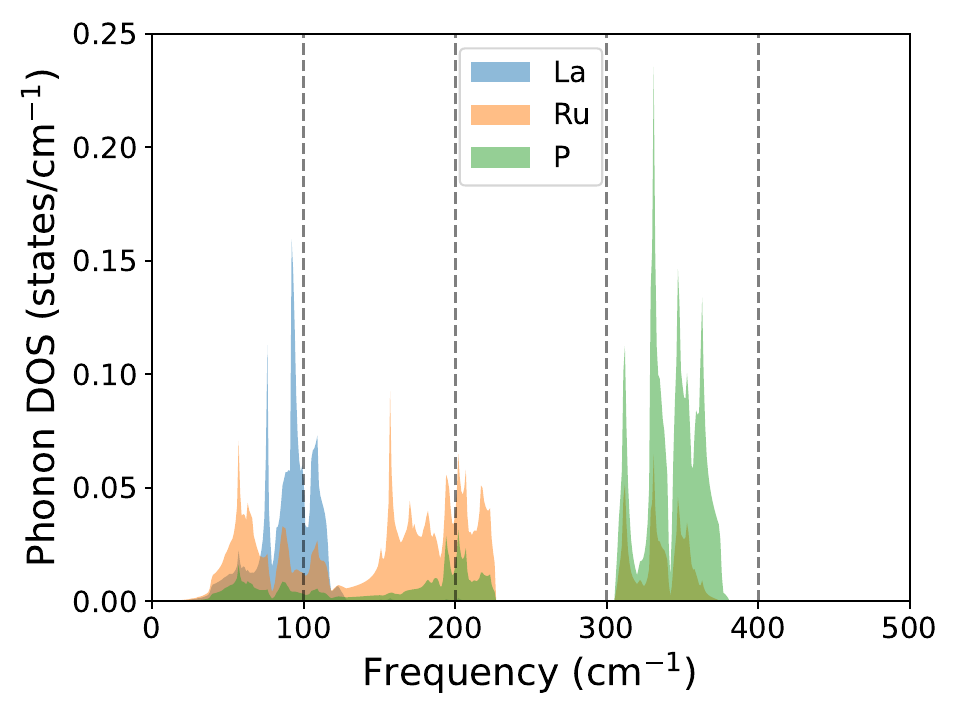}
  \includegraphics[width=0.195\hsize]{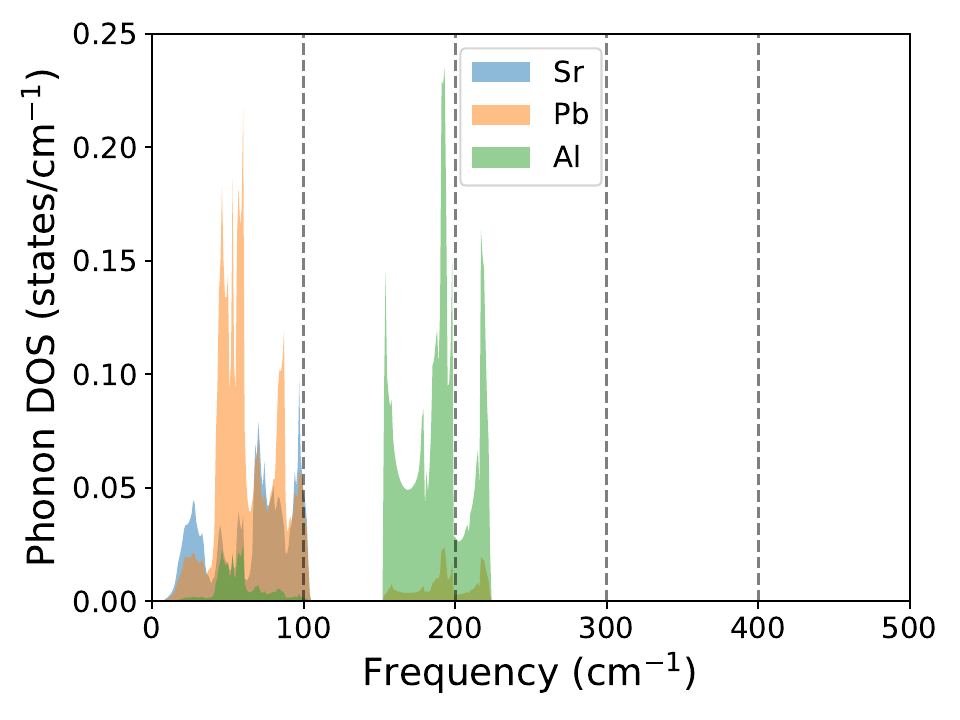}
  \includegraphics[width=0.195\hsize]{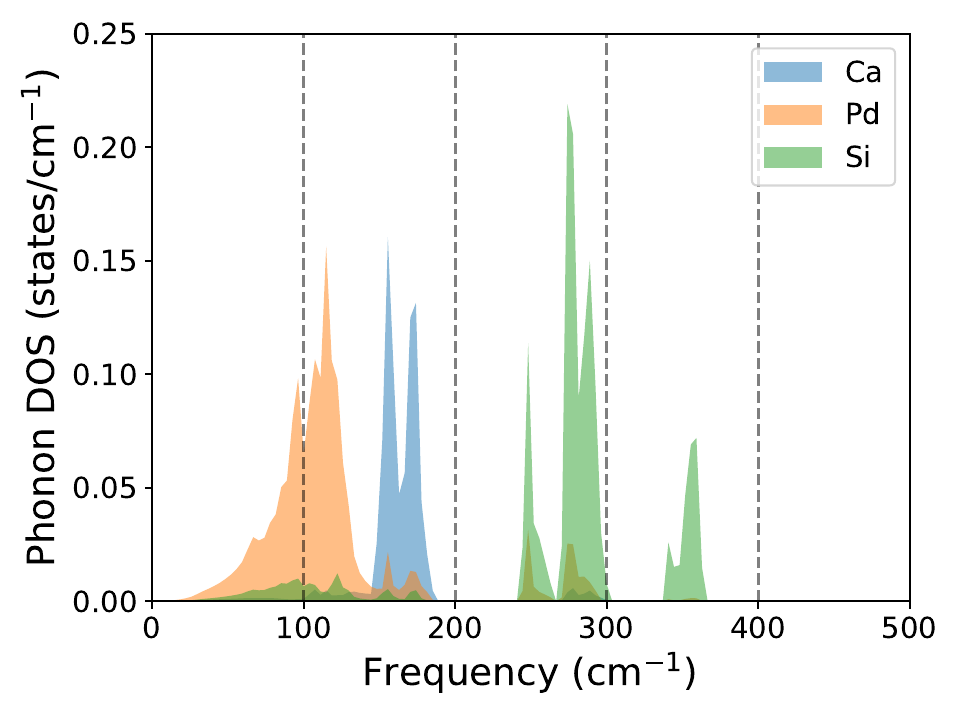}
  \includegraphics[width=0.195\hsize]{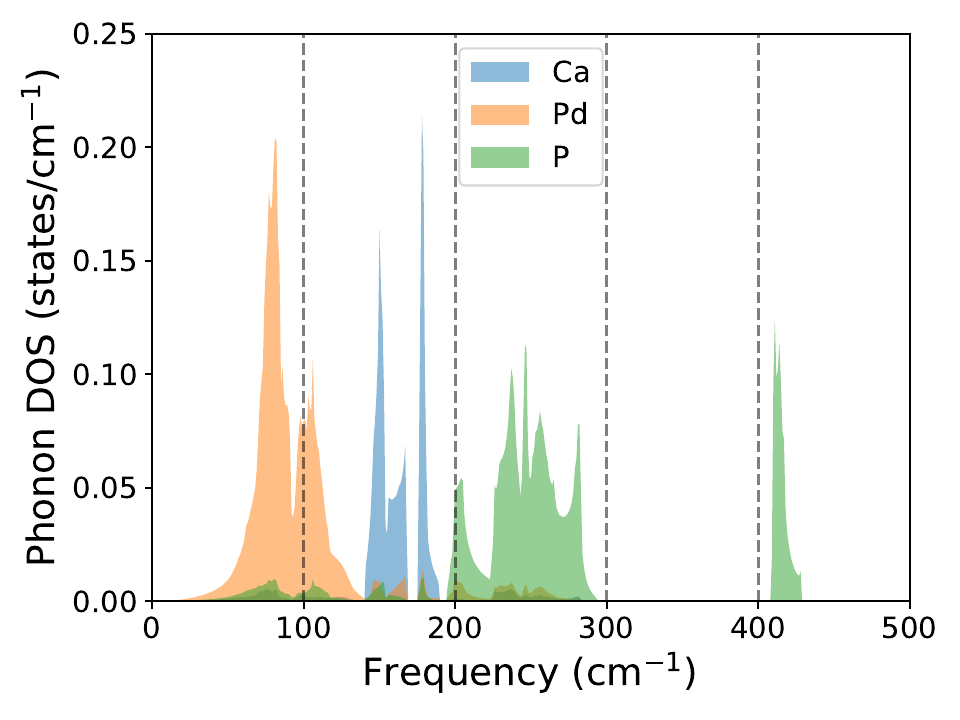}
  \\ 
  \includegraphics[width=0.195\hsize]{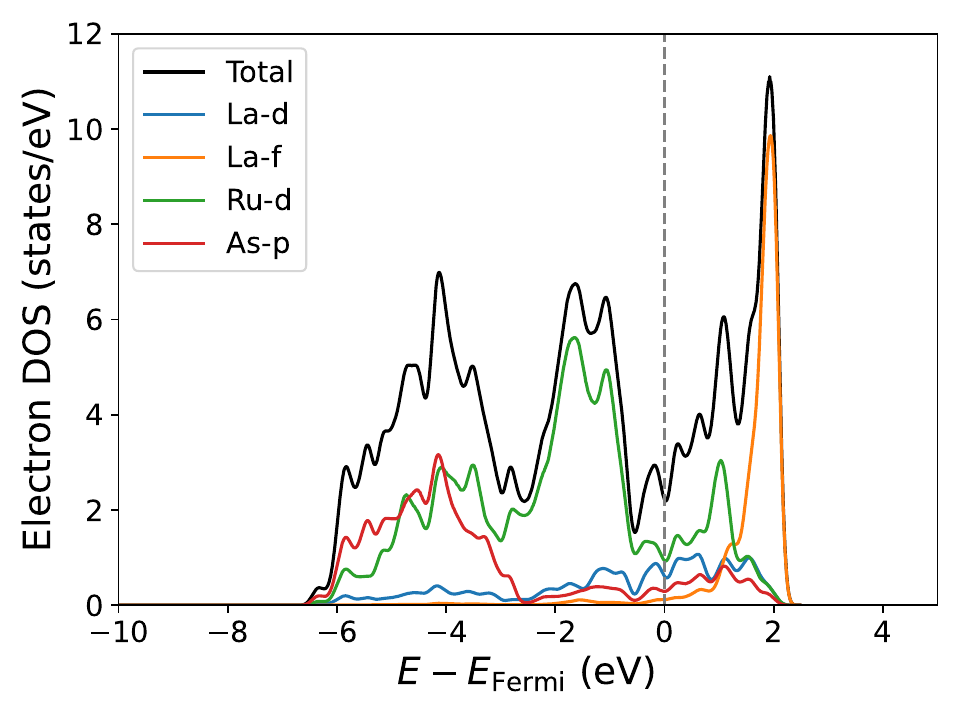}
  \includegraphics[width=0.195\hsize]{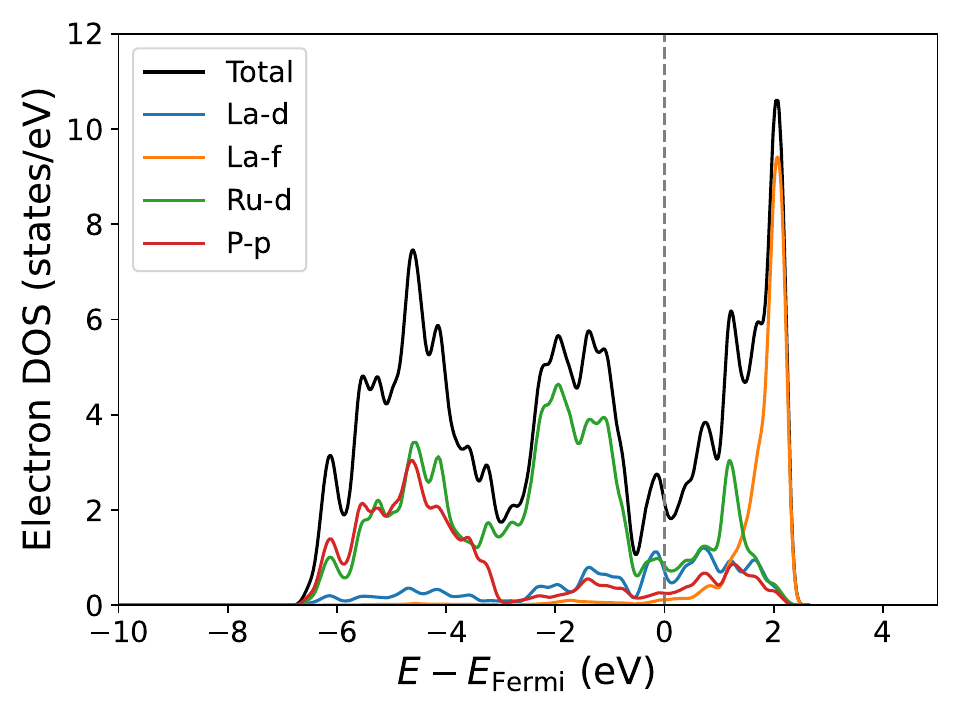}
  \includegraphics[width=0.195\hsize]{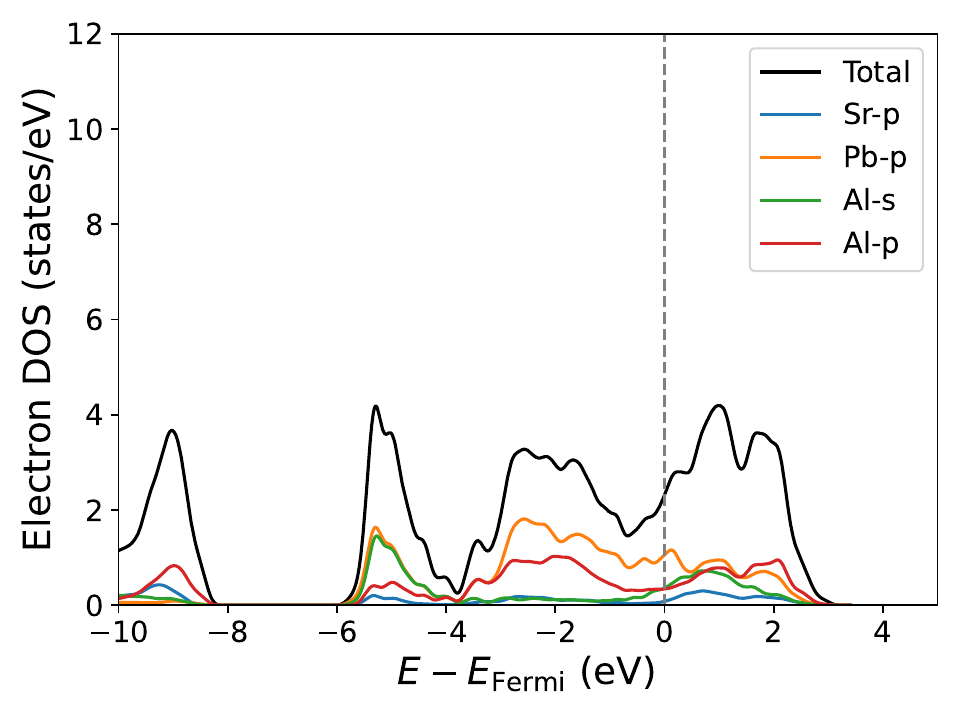}
  \includegraphics[width=0.195\hsize]{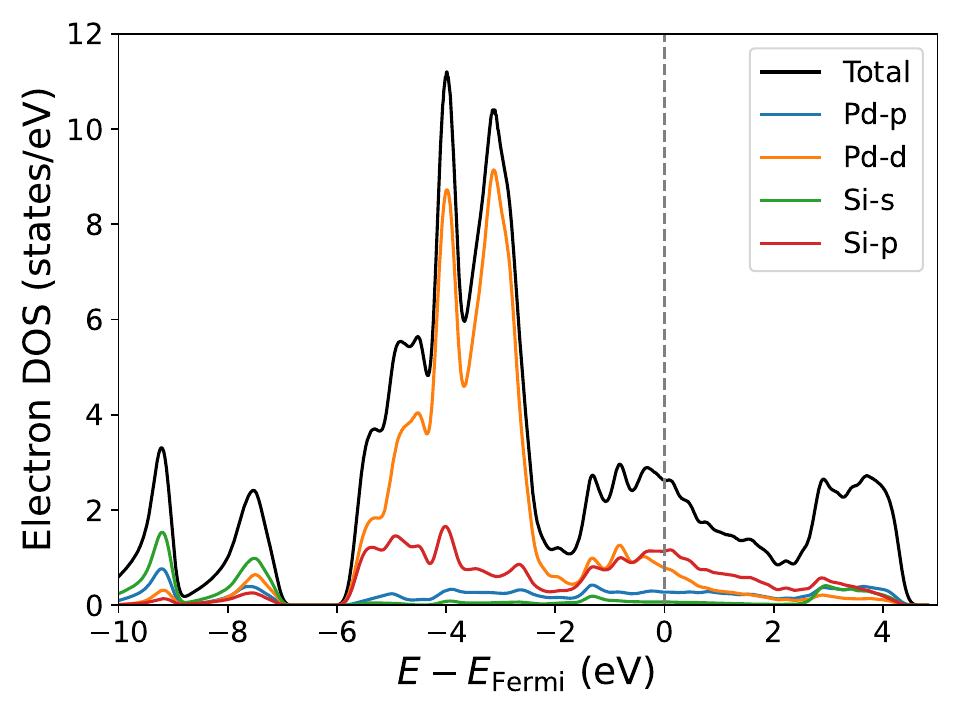}
  \includegraphics[width=0.195\hsize]{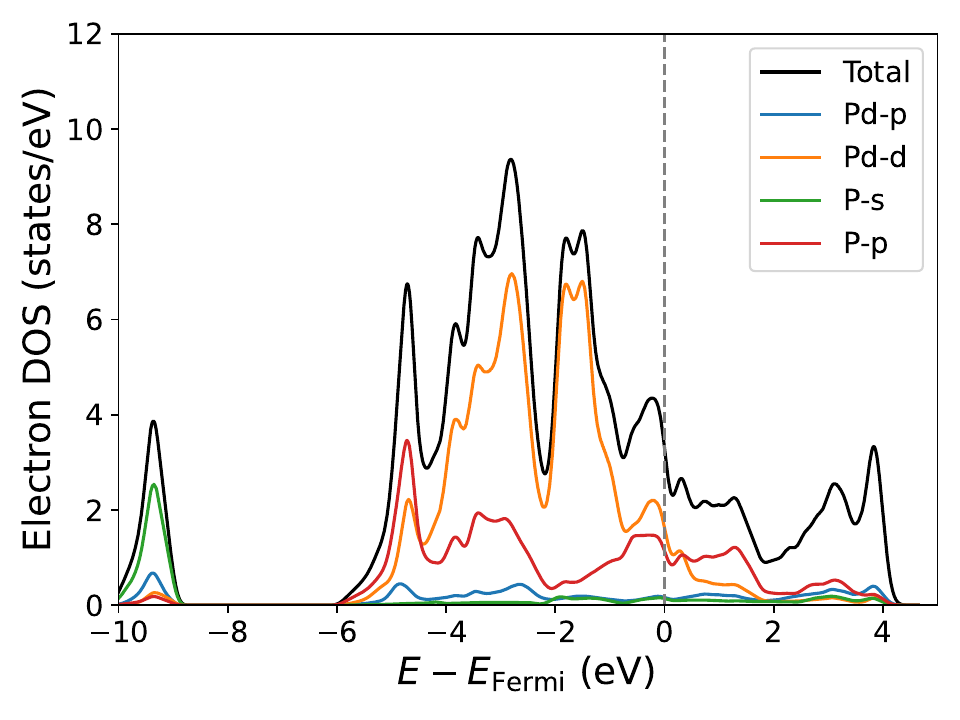}
  \\ 
  \includegraphics[width=0.195\hsize]{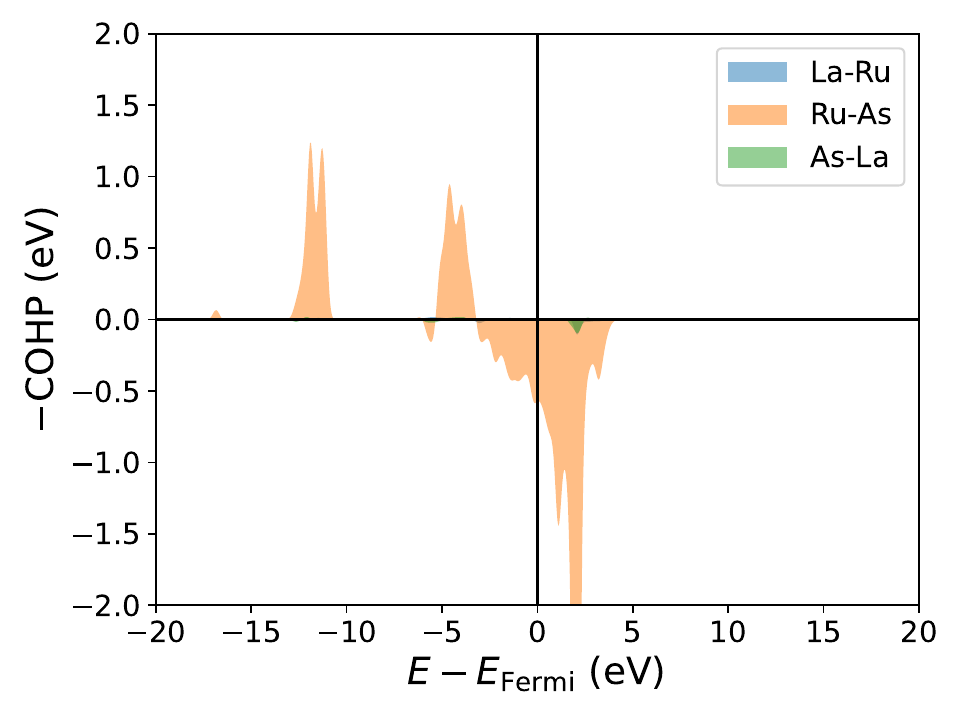}
  \includegraphics[width=0.195\hsize]{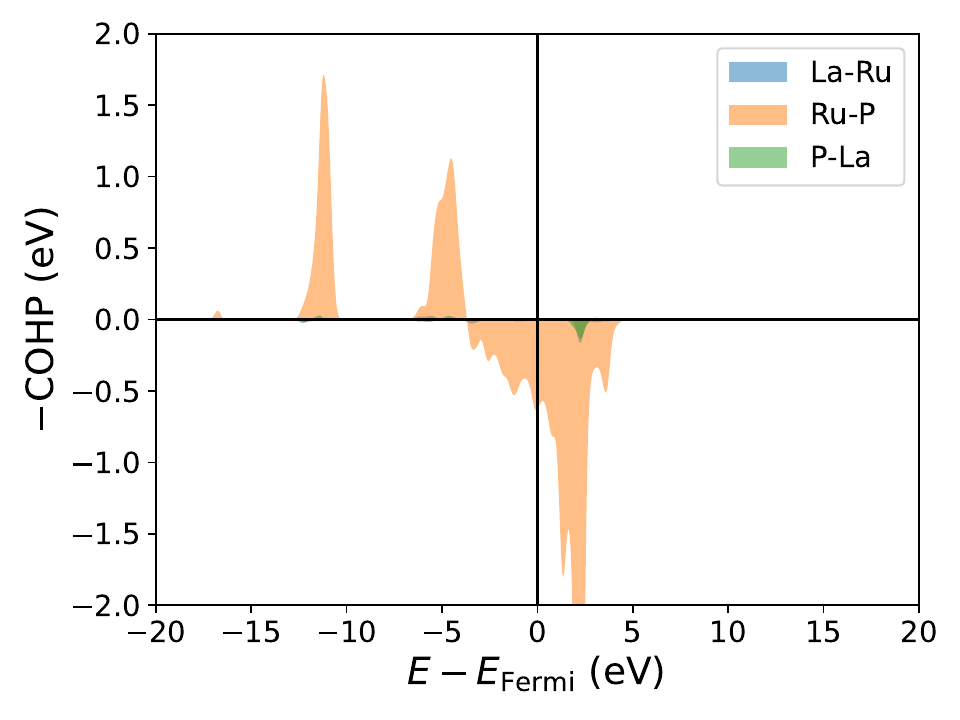}
  \includegraphics[width=0.195\hsize]{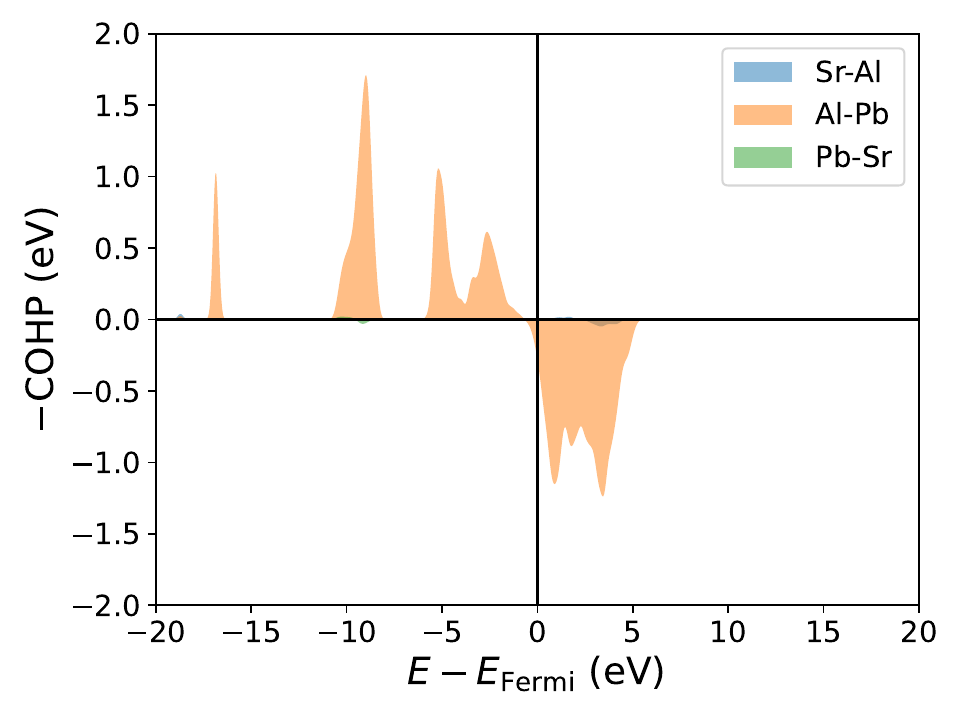}
  \includegraphics[width=0.195\hsize]{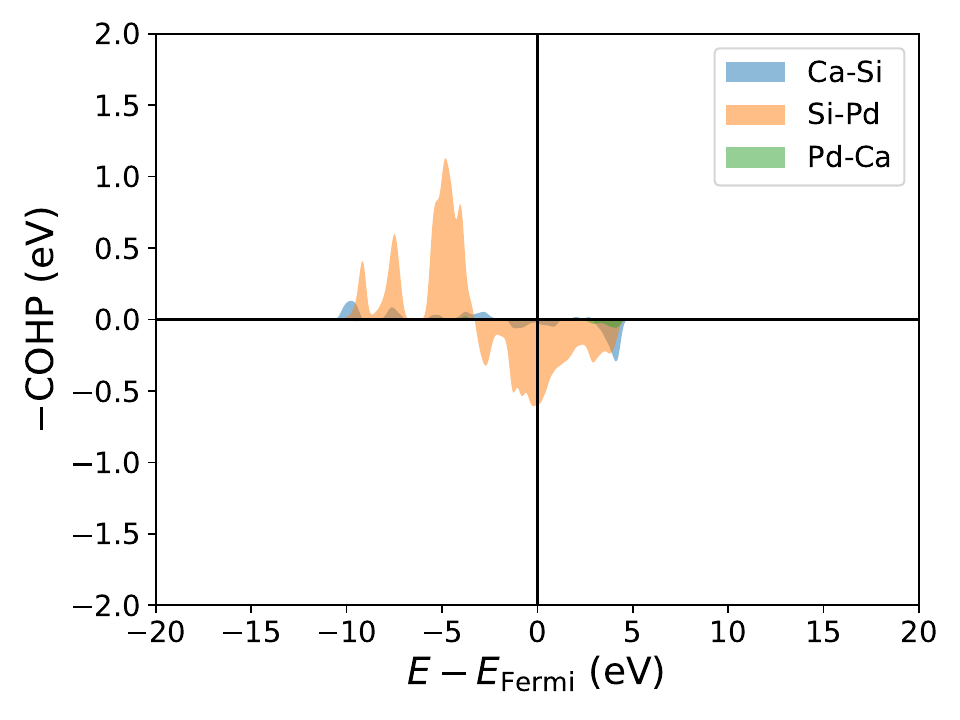}
  \includegraphics[width=0.195\hsize]{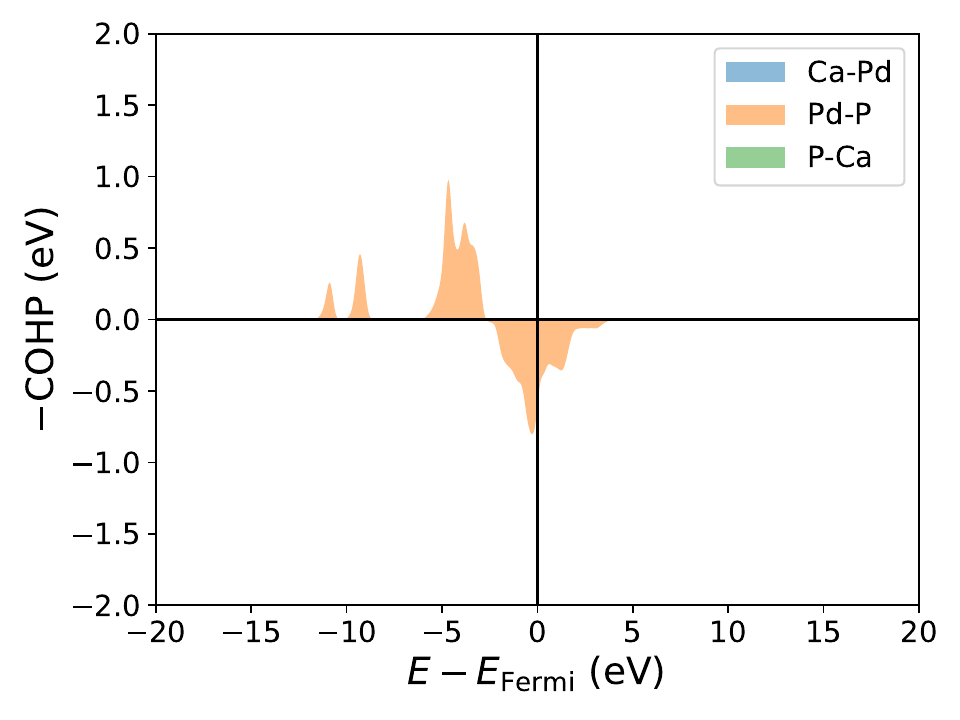}
  \\ 
  \begin{minipage}[t]{0.195\hsize}
    \includegraphics[width=\hsize]{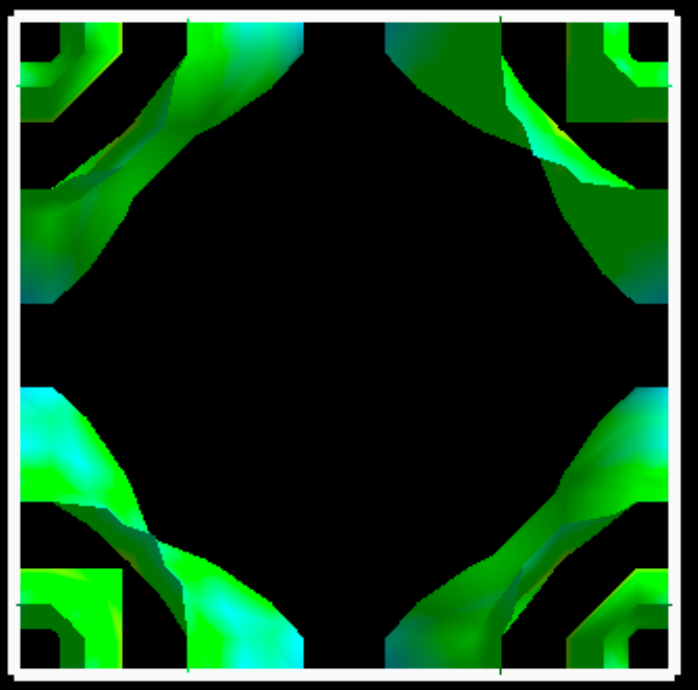}
    \includegraphics[width=\hsize]{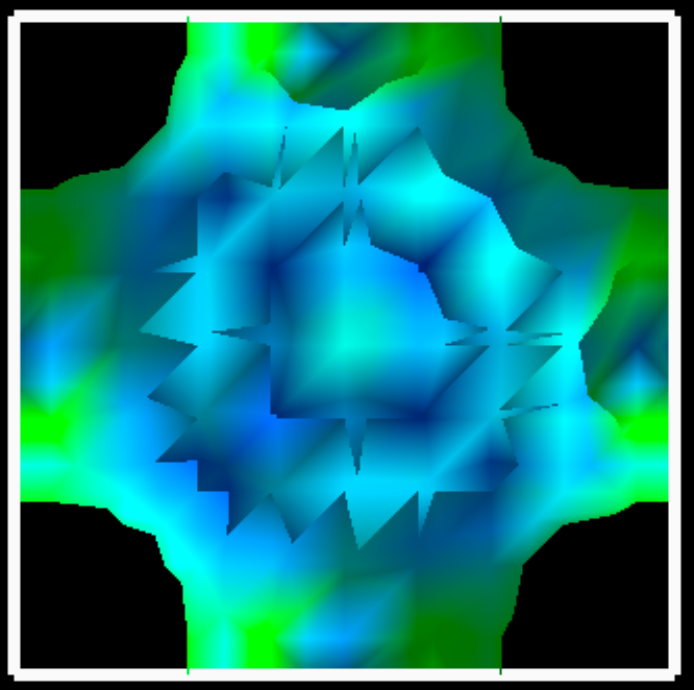}
  \end{minipage}
  \begin{minipage}[t]{0.195\hsize}
    \includegraphics[width=\hsize]{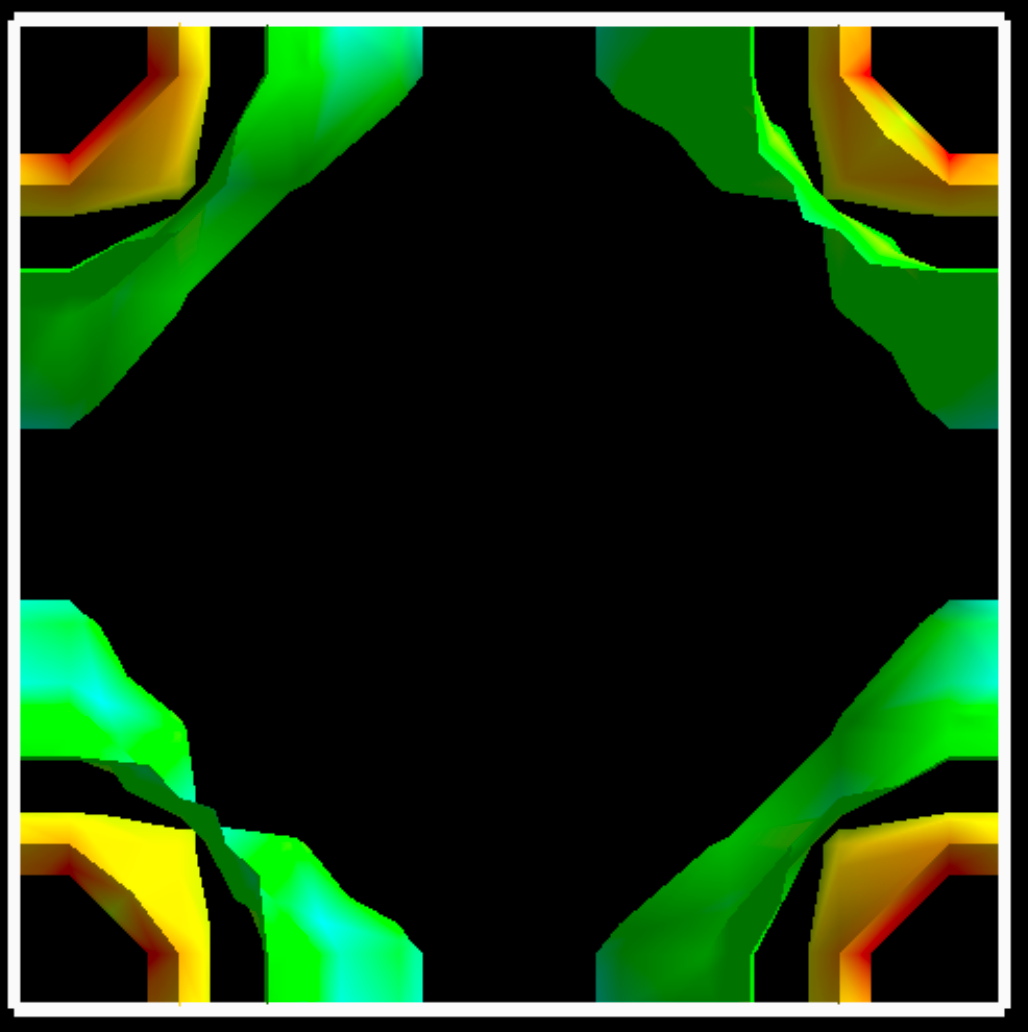}
    \includegraphics[width=\hsize]{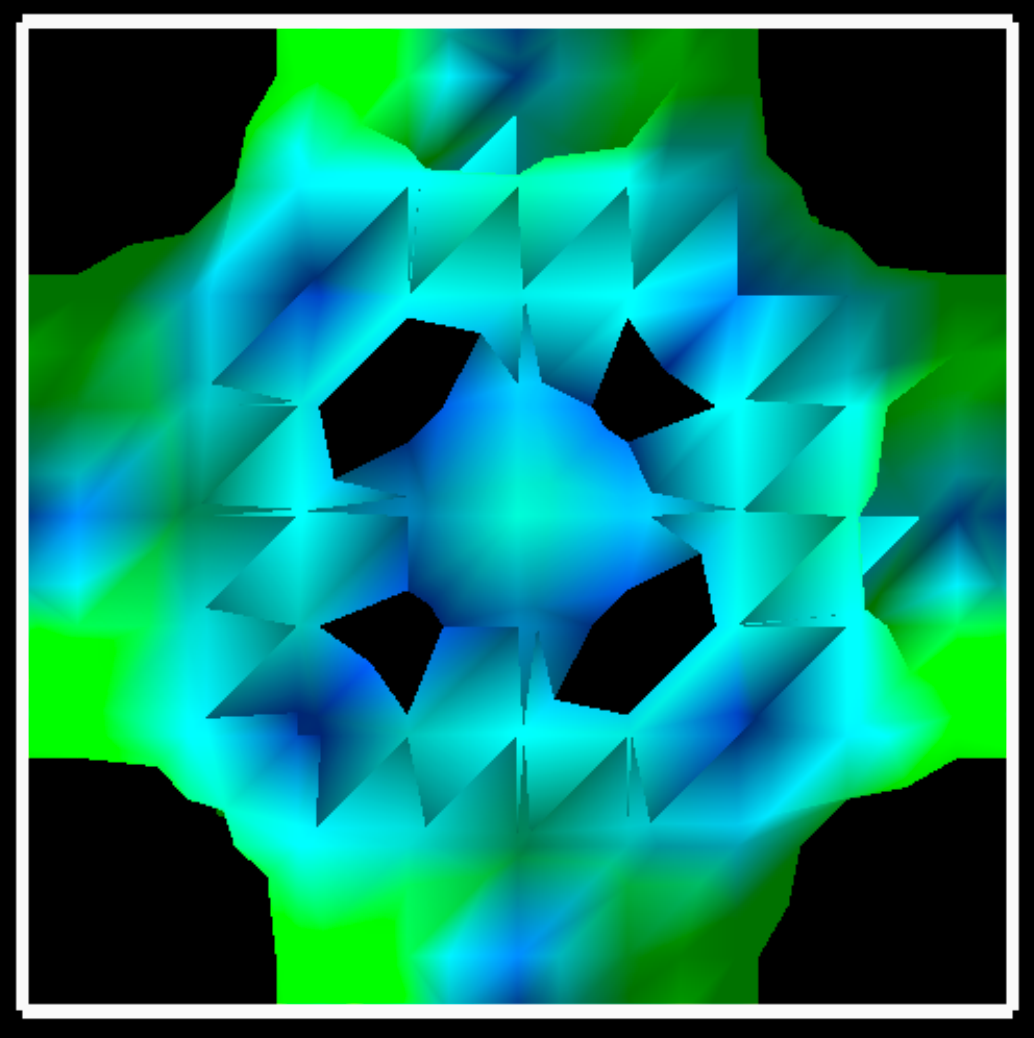}
  \end{minipage}
  \begin{minipage}[t]{0.195\hsize}
    \includegraphics[width=\hsize]{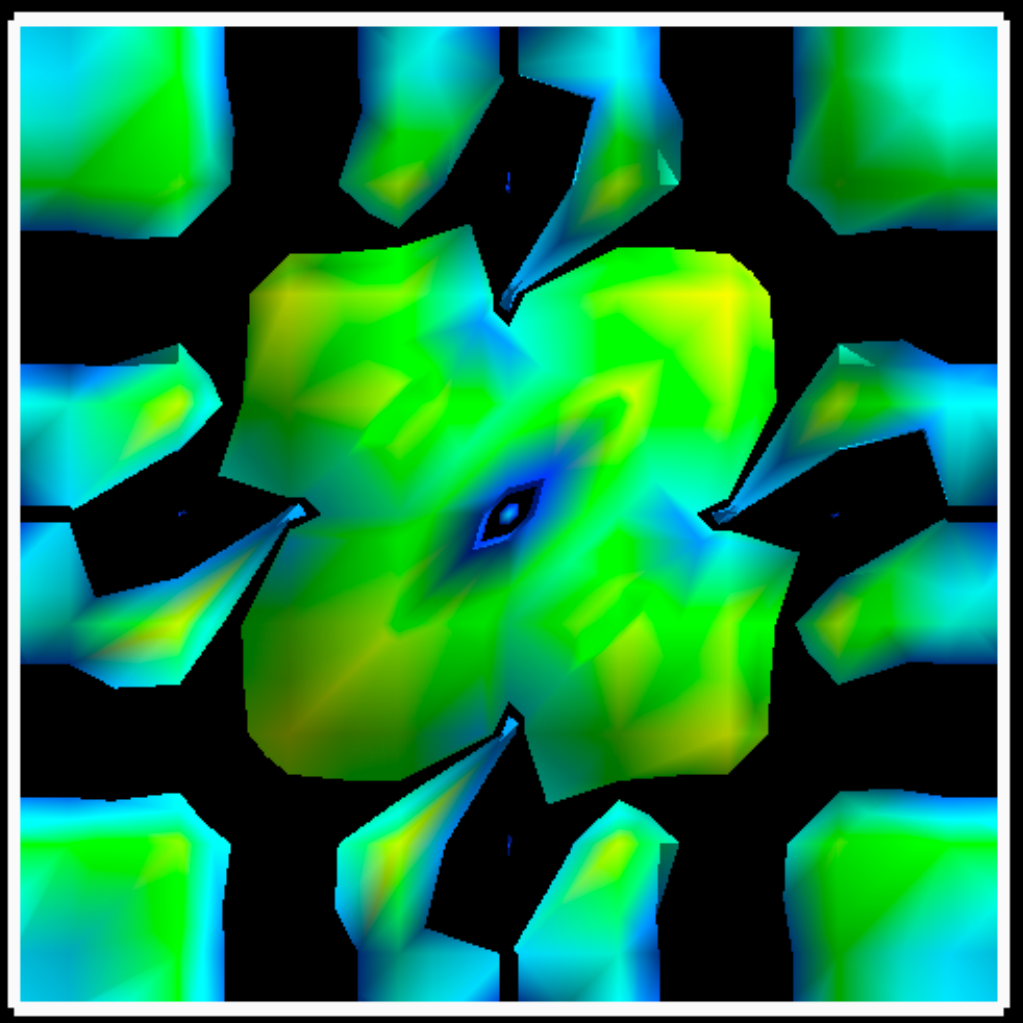}
  \end{minipage}
  \begin{minipage}[t]{0.195\hsize}
    \includegraphics[width=\hsize]{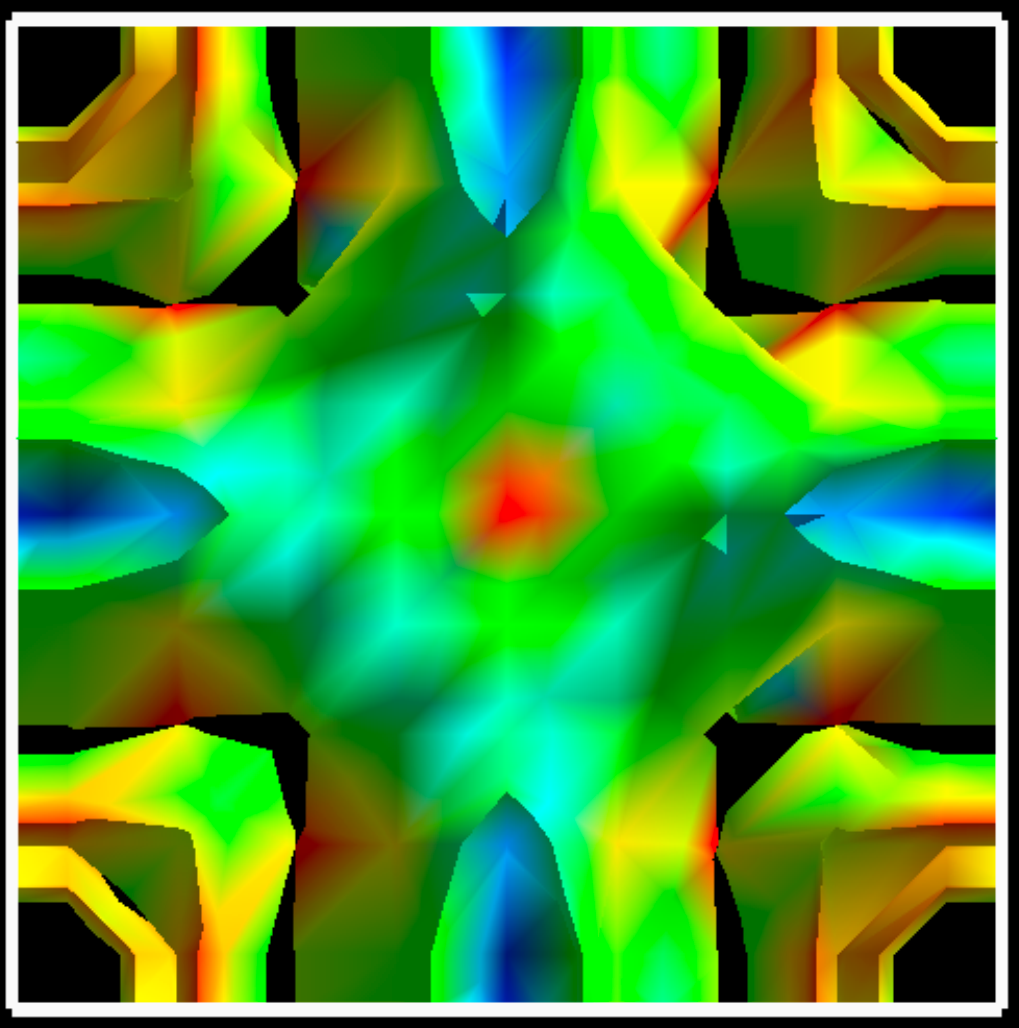}
  \end{minipage}
  \begin{minipage}[t]{0.195\hsize}
    \includegraphics[width=\hsize]{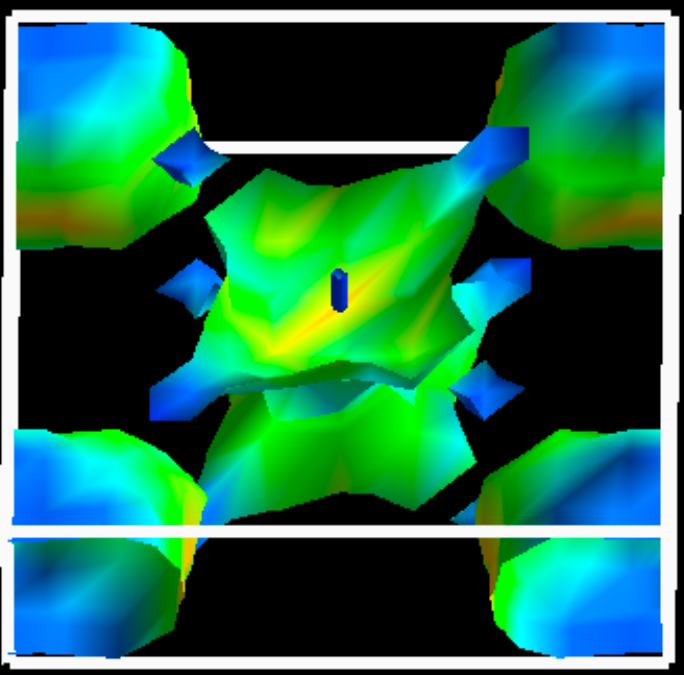}
  \end{minipage}
  \caption{
    \label{fig:elecPhonGraphs}\ghost{fig:elecPhonGraphs}
    \tadd{
      Comparison of the Eliashberg function $\alpha^2F(\omega)$, cumulative electron-phonon
      coupling constant $\lambda(\omega)$, phonon density of states (DOS), electronic DOS,
      crystal orbital Hamilton population (COHP), and Fermi surfaces
      for representative superconductors discussed in the text
      (LaRu$_2$As$_2$, LaRu$_2$P$_2$, SrPb$_2$Al$_2$, CaPd$_2$Si$_2$, CaPd$_2$P$_2$).
    }
  }
\end{figure*}

\vspace{2mm}
\tadd{
  Figure~\ref{fig:elecPhonGraphs} indicates the electron and phonon
  properties of notable compounds, based on which we discuss
  the superconducting mechanisms of each compound.
}

\vspace{2mm}
\tadd{
  LaRu$_2$As$_2$ and LaRu$_2$P$_2$, which exhibit particularly high {\tc},
  possess characteristically moderately two-dimensional electron pockets:
  their nesting could enhance the electron-phonon interactions \cite{2018Li_Broido}.
  Another contributing factor could be that strongly antibonding orbitals
  between Ru and As/P are located at the Fermi level:
  antibonding orbitals are more susceptible to geometric changes,
  and thus the electronic energy changes more significantly
  in response to lattice distortions,
  resulting in larger electron-phonon interactions.
  This is because, in antibonding orbitals, wavefunctions with opposite signs
  face each other at the bonding region, leading to strong overlap repulsion.
  By comparing the $\omega$-dependence of the cumulative $\lambda$ and phonon DOS,
  it is found that relatively low-frequency phonon modes, mainly originating
  from vibrations of the Ru-As/P layers, contribute to the increase in $\lambda$.
  Comparing LaRu$_2$As$_2$ and LaRu$_2$P$_2$,
  the Ru-As modes have lower frequencies than the Ru-P ones,
  which is considered to be a reason for the higher $\lambda$ in LaRu$_2$As$_2$.
  In addition to the heavier mass of As, a comparison of the integrated COHP (ICOHP)
  also predicts that the Ru-As bond is weaker
  (Ru-As's -ICOHP=2.32~eV, Ru-P's -ICOHP=2.48~eV),
  leading to the softer Ru-As modes.
  Furthermore, the two-dimensionality of the electron pockets appears
  to be more pronounced in LaRu$_2$As$_2$,
  which may also contribute to its higher {\tc}.
}

\vspace{2mm}
\tadd{
  Among the newly identified superconductors, SrPb$_2$Al$_2$ exhibits the highest predicted {\tc}.
  In this compound, the Fermi surface topology is markedly different from that of LaRu$_2$As(P)$_2$,
  being three-dimensional and showing no evidence of nesting.
  Furthermore, COHP analysis reveals that the antibonding character at the Fermi level is also small.
  On the other hand, the presence of very low-frequency phonon modes originating from the heavy Pb and Sr atoms leads to
  a significant increase in $\lambda$ in that frequency range, resulting in a $\lambda$ value comparable to that of LaRu$_2$P$_2$.
  Regarding {\omegaln}, the molecular weight of SrPb$_2$Al$_2$ (555.98~g/mol) is heavier than
  that of LaRu$_2$P$_2$ (402.99~g/mol), which lowers {\omegaln} and consequently results in a lower {\tc}.
}

\vspace{2mm}
On the other hand, in CaPd$_2$Si$_2$, which shows the second-highest {\tccalc}
among the unreported compounds,
$\lambda$ is relatively modest ($\lambda = 0.46$), while {\omegaln} is notably large.
As shown in Eq.~\eqref{eq:allen-dynes}, within the BCS framework, higher {\tc} values are
achieved through larger $\lambda$ and higher {\omegaln}.
These two factors are generally known to be in a trade-off relationship,
and SrPb$_2$Al$_2$ (CaPd$_2$Si$_2$) can be considered a system
that gains {\tc} predominantly through $\lambda$ ({\omegaln}).

\vspace{2mm}
The value of {\omegaln} for CaPd$_2$Si$_2$ is significantly
higher than that of CaPd$_2$P$_2$ ({\omegaln} = 165.06~K),
which has the closest composition among the materials studied.
\footnote{
  The only difference between CaPd$_2$Si$_2$ and CaPd$_2$P$_2$ lies in the Si/P site.
  The atomic number of Si is 14, whereas that of P is 15.
}
\tadd{Comparing the cumulative $\lambda$ and phonon DOS,}
while {\omegaln} is primarily influenced by the distribution of phonon DOS in the low-frequency region,
the partial phonon DOS of Pd, dominant in the lowest frequency range, is shifted to higher frequencies
in CaPd$_2$Si$_2$ relative to CaPd$_2$P$_2$.
This shift likely contributes to the higher {\omegaln} observed in CaPd$_2$Si$_2$.

\vspace{2mm}
Such a shift suggests that Pd is subject to a harder potential in CaPd$_2$Si$_2$,
indicating stronger bonding.
One plausible explanation for this difference is the variation in electronegativity between Si and P.
As the nearest-neighbor atoms to Pd, their electronegativities are Pd: 2.20, Si: 1.90, and P: 2.19~\cite{1960PAULING}.
The larger electronegativity difference between Pd and Si, compared to that between Pd and P,
implies that Pd in CaPd$_2$Si$_2$ is subject to a harder potential from Si,
resulting in harder vibrational modes.

\section{Conclusions}\label{sec.conc}\ghost{sec.conc}
In this study, we systematically explored potential BCS-type superconductors
among {\tcs}-type compounds using {\abinitio} {\tc} calculations.
Starting from {\NumDatabase} candidate compounds in the ICSD \cite{1983BER},
we excluded those likely to hinder BCS-type superconductivity:
namely, compounds containing magnetic elements (Cr, Mn, Fe, Co, or Ni)
and those with $f$-electron elements that may exhibit heavy-fermion behavior.
We also removed compounds with imaginary phonon modes indicative of dynamic instability,
resulting in {\NumStable} stable candidates.
Among these, {\NumReported} compounds have experimentally reported {\tc} values,
which were used to validate our {\abinitio} {\tc} calculations.
The calculated {\tc} values showed good overall agreement with experiments,
except for LaRu$_2$Si$_2$ ({\tcexpt} = 3.5~K) and LuRu$_2$Si$_2$ ({\tcexpt} = 2.4~K),
both of which were predicted to be nearly non-superconducting.
These discrepancies may stem from spin-fluctuation-mediated superconductivity,
as experimental studies suggest proximity to a Stoner instability in these materials.
Excluding these two cases, the correlation coefficient between calculated and experimental $T_c$ improves to 0.86,
indicating high predictive reliability for BCS-type systems.
We then applied the same method to the {\NumUnreported} compounds lacking experimental
data and identified several new BCS-type superconductor candidates,
including SrPb$_2$Al$_2$ ({\tccalc} = 2.2~K).

\section*{Acknowledgments}
The computations in this work have been performed using the facilities of the Research Center for Advanced Computing Infrastructure (RCACI) at JAIST.
T.I. appreciate the support from JSPS KAKENHI Grant Number 24K17618 and JSPS Overseas Research Fellowships.
R.M. is grateful for financial support from
MEXT-KAKENHI (JP19H04692 and JP16KK0097),
from FLAGSHIP2020 (project nos. hp190169 and hp190167 at K-computer),
from Toyota Motor Corporation,
from the Air Force Office of Scientific Research
(AFOSR-AOARD/FA2386-17-1-4049;FA2386-19-1-4015),
and from JSPS Bilateral Joint Projects (with India DST).
K.H. is grateful for financial support from
the HPCI System Research Project (Project ID: hp190169) and
MEXT-KAKENHI (JP16H06439, JP17K17762, JP19K05029, and JP19H05169).

\bibliography{references}

\begin{thebibliography}{61}%
\makeatletter
\providecommand \@ifxundefined [1]{%
 \@ifx{#1\undefined}
}%
\providecommand \@ifnum [1]{%
 \ifnum #1\expandafter \@firstoftwo
 \else \expandafter \@secondoftwo
 \fi
}%
\providecommand \@ifx [1]{%
 \ifx #1\expandafter \@firstoftwo
 \else \expandafter \@secondoftwo
 \fi
}%
\providecommand \natexlab [1]{#1}%
\providecommand \enquote  [1]{``#1''}%
\providecommand \bibnamefont  [1]{#1}%
\providecommand \bibfnamefont [1]{#1}%
\providecommand \citenamefont [1]{#1}%
\providecommand \href@noop [0]{\@secondoftwo}%
\providecommand \href [0]{\begingroup \@sanitize@url \@href}%
\providecommand \@href[1]{\@@startlink{#1}\@@href}%
\providecommand \@@href[1]{\endgroup#1\@@endlink}%
\providecommand \@sanitize@url [0]{\catcode `\\12\catcode `\$12\catcode
  `\&12\catcode `\#12\catcode `\^12\catcode `\_12\catcode `\%12\relax}%
\providecommand \@@startlink[1]{}%
\providecommand \@@endlink[0]{}%
\providecommand \url  [0]{\begingroup\@sanitize@url \@url }%
\providecommand \@url [1]{\endgroup\@href {#1}{\urlprefix }}%
\providecommand \urlprefix  [0]{URL }%
\providecommand \Eprint [0]{\href }%
\providecommand \doibase [0]{https://doi.org/}%
\providecommand \selectlanguage [0]{\@gobble}%
\providecommand \bibinfo  [0]{\@secondoftwo}%
\providecommand \bibfield  [0]{\@secondoftwo}%
\providecommand \translation [1]{[#1]}%
\providecommand \BibitemOpen [0]{}%
\providecommand \bibitemStop [0]{}%
\providecommand \bibitemNoStop [0]{.\EOS\space}%
\providecommand \EOS [0]{\spacefactor3000\relax}%
\providecommand \BibitemShut  [1]{\csname bibitem#1\endcsname}%
\let\auto@bib@innerbib\@empty
\bibitem [{\citenamefont {Zhou}\ \emph {et~al.}(2021)\citenamefont {Zhou},
  \citenamefont {Lee}, \citenamefont {Imada}, \citenamefont {Trivedi},
  \citenamefont {Phillips}, \citenamefont {Kee}, \citenamefont
  {T{\"o}rm{\"a}},\ and\ \citenamefont {Eremets}}]{2021Xingjiang_Eremets}%
  \BibitemOpen
  \bibfield  {author} {\bibinfo {author} {\bibfnamefont {X.}~\bibnamefont
  {Zhou}}, \bibinfo {author} {\bibfnamefont {W.-S.}\ \bibnamefont {Lee}},
  \bibinfo {author} {\bibfnamefont {M.}~\bibnamefont {Imada}}, \bibinfo
  {author} {\bibfnamefont {N.}~\bibnamefont {Trivedi}}, \bibinfo {author}
  {\bibfnamefont {P.}~\bibnamefont {Phillips}}, \bibinfo {author}
  {\bibfnamefont {H.-Y.}\ \bibnamefont {Kee}}, \bibinfo {author} {\bibfnamefont
  {P.}~\bibnamefont {T{\"o}rm{\"a}}},\ and\ \bibinfo {author} {\bibfnamefont
  {M.}~\bibnamefont {Eremets}},\ }\bibfield  {title} {\bibinfo {title}
  {High-temperature superconductivity},\ }\href
  {https://doi.org/10.1038/s42254-021-00324-3} {\bibfield  {journal} {\bibinfo
  {journal} {Nature Reviews Physics}\ }\textbf {\bibinfo {volume} {3}},\
  \bibinfo {pages} {462} (\bibinfo {year} {2021})}\BibitemShut {NoStop}%
\bibitem [{\citenamefont {Hayden}\ and\ \citenamefont
  {Tranquada}(2024)}]{2024Hayden_Tranquada}%
  \BibitemOpen
  \bibfield  {author} {\bibinfo {author} {\bibfnamefont {S.~M.}\ \bibnamefont
  {Hayden}}\ and\ \bibinfo {author} {\bibfnamefont {J.~M.}\ \bibnamefont
  {Tranquada}},\ }\bibfield  {title} {\bibinfo {title} {Charge correlations in
  cuprate superconductors},\ }\href
  {https://doi.org/https://doi.org/10.1146/annurev-conmatphys-032922-094430}
  {\bibfield  {journal} {\bibinfo  {journal} {Annual Review of Condensed Matter
  Physics}\ }\textbf {\bibinfo {volume} {15}},\ \bibinfo {pages} {215}
  (\bibinfo {year} {2024})}\BibitemShut {NoStop}%
\bibitem [{\citenamefont {Mark}\ \emph {et~al.}(2022)\citenamefont {Mark},
  \citenamefont {Campuzano},\ and\ \citenamefont
  {Hemley}}]{2022Alexander_Hemley}%
  \BibitemOpen
  \bibfield  {author} {\bibinfo {author} {\bibfnamefont {A.~C.}\ \bibnamefont
  {Mark}}, \bibinfo {author} {\bibfnamefont {J.~C.}\ \bibnamefont
  {Campuzano}},\ and\ \bibinfo {author} {\bibfnamefont {R.~J.}\ \bibnamefont
  {Hemley}},\ }\bibfield  {title} {\bibinfo {title} {Progress and prospects for
  cuprate high temperature superconductors under pressure},\ }\href
  {https://doi.org/10.1080/08957959.2022.2059366} {\bibfield  {journal}
  {\bibinfo  {journal} {High Pressure Research}\ }\textbf {\bibinfo {volume}
  {42}},\ \bibinfo {pages} {137} (\bibinfo {year} {2022})},\ \Eprint
  {https://arxiv.org/abs/https://doi.org/10.1080/08957959.2022.2059366}
  {https://doi.org/10.1080/08957959.2022.2059366} \BibitemShut {NoStop}%
\bibitem [{\citenamefont {Boebinger}\ \emph {et~al.}(2024)\citenamefont
  {Boebinger}, \citenamefont {Chubukov}, \citenamefont {Fisher}, \citenamefont
  {Grosche}, \citenamefont {Hirschfeld}, \citenamefont {Julian}, \citenamefont
  {Keimer}, \citenamefont {Kivelson}, \citenamefont {Mackenzie}, \citenamefont
  {Maeno}, \citenamefont {Orenstein}, \citenamefont {Ramshaw}, \citenamefont
  {Sachdev}, \citenamefont {Schmalian},\ and\ \citenamefont
  {Vojta}}]{2024Boebinger_Vojta}%
  \BibitemOpen
  \bibfield  {author} {\bibinfo {author} {\bibfnamefont {G.~S.}\ \bibnamefont
  {Boebinger}}, \bibinfo {author} {\bibfnamefont {A.~V.}\ \bibnamefont
  {Chubukov}}, \bibinfo {author} {\bibfnamefont {I.~R.}\ \bibnamefont
  {Fisher}}, \bibinfo {author} {\bibfnamefont {F.~M.}\ \bibnamefont {Grosche}},
  \bibinfo {author} {\bibfnamefont {P.~J.}\ \bibnamefont {Hirschfeld}},
  \bibinfo {author} {\bibfnamefont {S.~R.}\ \bibnamefont {Julian}}, \bibinfo
  {author} {\bibfnamefont {B.}~\bibnamefont {Keimer}}, \bibinfo {author}
  {\bibfnamefont {S.~A.}\ \bibnamefont {Kivelson}}, \bibinfo {author}
  {\bibfnamefont {A.~P.}\ \bibnamefont {Mackenzie}}, \bibinfo {author}
  {\bibfnamefont {Y.}~\bibnamefont {Maeno}}, \bibinfo {author} {\bibfnamefont
  {J.}~\bibnamefont {Orenstein}}, \bibinfo {author} {\bibfnamefont {B.~J.}\
  \bibnamefont {Ramshaw}}, \bibinfo {author} {\bibfnamefont {S.}~\bibnamefont
  {Sachdev}}, \bibinfo {author} {\bibfnamefont {J.}~\bibnamefont {Schmalian}},\
  and\ \bibinfo {author} {\bibfnamefont {M.}~\bibnamefont {Vojta}},\ }\bibfield
   {title} {\bibinfo {title} {Hydride superconductivity is here to stay},\
  }\href {https://doi.org/10.1038/s42254-024-00794-1} {\bibfield  {journal}
  {\bibinfo  {journal} {Nature Reviews Physics}\ }\textbf {\bibinfo {volume}
  {7}},\ \bibinfo {pages} {2} (\bibinfo {year} {2024})}\BibitemShut {NoStop}%
\bibitem [{\citenamefont {Nekrasov}\ and\ \citenamefont
  {Ovchinnikov}(2022)}]{2022Nekrasov_Sergei}%
  \BibitemOpen
  \bibfield  {author} {\bibinfo {author} {\bibfnamefont {I.}~\bibnamefont
  {Nekrasov}}\ and\ \bibinfo {author} {\bibfnamefont {S.}~\bibnamefont
  {Ovchinnikov}},\ }\bibfield  {title} {\bibinfo {title} {Hydrides under high
  pressure},\ }\href {https://doi.org/10.1007/s10948-021-06087-3} {\bibfield
  {journal} {\bibinfo  {journal} {Journal of Superconductivity and Novel
  Magnetism}\ }\textbf {\bibinfo {volume} {35}},\ \bibinfo {pages} {959}
  (\bibinfo {year} {2022})}\BibitemShut {NoStop}%
\bibitem [{\citenamefont {Sun}\ \emph {et~al.}(2023)\citenamefont {Sun},
  \citenamefont {Zhong}, \citenamefont {Liu},\ and\ \citenamefont
  {Ma}}]{2023Sun_Yanming}%
  \BibitemOpen
  \bibfield  {author} {\bibinfo {author} {\bibfnamefont {Y.}~\bibnamefont
  {Sun}}, \bibinfo {author} {\bibfnamefont {X.}~\bibnamefont {Zhong}}, \bibinfo
  {author} {\bibfnamefont {H.}~\bibnamefont {Liu}},\ and\ \bibinfo {author}
  {\bibfnamefont {Y.}~\bibnamefont {Ma}},\ }\bibfield  {title} {\bibinfo
  {title} {Clathrate metal superhydrides under high-pressure conditions:
  enroute to room-temperature superconductivity},\ }\href
  {https://doi.org/10.1093/nsr/nwad270} {\bibfield  {journal} {\bibinfo
  {journal} {National Science Review}\ }\textbf {\bibinfo {volume} {11}},\
  \bibinfo {pages} {nwad270} (\bibinfo {year} {2023})},\ \Eprint
  {https://arxiv.org/abs/https://academic.oup.com/nsr/article-pdf/11/7/nwad270/58227797/nwad270.pdf}
  {https://academic.oup.com/nsr/article-pdf/11/7/nwad270/58227797/nwad270.pdf}
  \BibitemShut {NoStop}%
\bibitem [{\citenamefont {Zhao}\ \emph {et~al.}(2023)\citenamefont {Zhao},
  \citenamefont {Huang}, \citenamefont {Zhang}, \citenamefont {Chen},
  \citenamefont {Du}, \citenamefont {Duan},\ and\ \citenamefont
  {Cui}}]{2023Zhao_Cui}%
  \BibitemOpen
  \bibfield  {author} {\bibinfo {author} {\bibfnamefont {W.}~\bibnamefont
  {Zhao}}, \bibinfo {author} {\bibfnamefont {X.}~\bibnamefont {Huang}},
  \bibinfo {author} {\bibfnamefont {Z.}~\bibnamefont {Zhang}}, \bibinfo
  {author} {\bibfnamefont {S.}~\bibnamefont {Chen}}, \bibinfo {author}
  {\bibfnamefont {M.}~\bibnamefont {Du}}, \bibinfo {author} {\bibfnamefont
  {D.}~\bibnamefont {Duan}},\ and\ \bibinfo {author} {\bibfnamefont
  {T.}~\bibnamefont {Cui}},\ }\bibfield  {title} {\bibinfo {title}
  {Superconducting ternary hydrides: progress and challenges},\ }\href
  {https://doi.org/10.1093/nsr/nwad307} {\bibfield  {journal} {\bibinfo
  {journal} {National Science Review}\ }\textbf {\bibinfo {volume} {11}},\
  \bibinfo {pages} {nwad307} (\bibinfo {year} {2023})},\ \Eprint
  {https://arxiv.org/abs/https://academic.oup.com/nsr/article-pdf/11/7/nwad307/58227815/nwad307.pdf}
  {https://academic.oup.com/nsr/article-pdf/11/7/nwad307/58227815/nwad307.pdf}
  \BibitemShut {NoStop}%
\bibitem [{\citenamefont {Allen}\ and\ \citenamefont {Dynes}(1975)}]{1975ALL}%
  \BibitemOpen
  \bibfield  {author} {\bibinfo {author} {\bibfnamefont {P.~B.}\ \bibnamefont
  {Allen}}\ and\ \bibinfo {author} {\bibfnamefont {R.~C.}\ \bibnamefont
  {Dynes}},\ }\bibfield  {title} {\bibinfo {title} {Transition temperature of
  strong-coupled superconductors reanalyzed},\ }\href@noop {} {\bibfield
  {journal} {\bibinfo  {journal} {Phys. Rev. B}\ }\textbf {\bibinfo {volume}
  {12}},\ \bibinfo {pages} {905} (\bibinfo {year} {1975})}\BibitemShut
  {NoStop}%
\bibitem [{\citenamefont {Fernandes}\ \emph {et~al.}(2022)\citenamefont
  {Fernandes}, \citenamefont {Coldea}, \citenamefont {Ding}, \citenamefont
  {Fisher}, \citenamefont {Hirschfeld},\ and\ \citenamefont
  {Kotliar}}]{2022Fernandes_Kotliar}%
  \BibitemOpen
  \bibfield  {author} {\bibinfo {author} {\bibfnamefont {R.~M.}\ \bibnamefont
  {Fernandes}}, \bibinfo {author} {\bibfnamefont {A.~I.}\ \bibnamefont
  {Coldea}}, \bibinfo {author} {\bibfnamefont {H.}~\bibnamefont {Ding}},
  \bibinfo {author} {\bibfnamefont {I.~R.}\ \bibnamefont {Fisher}}, \bibinfo
  {author} {\bibfnamefont {P.~J.}\ \bibnamefont {Hirschfeld}},\ and\ \bibinfo
  {author} {\bibfnamefont {G.}~\bibnamefont {Kotliar}},\ }\bibfield  {title}
  {\bibinfo {title} {Iron pnictides and chalcogenides: a new paradigm for
  superconductivity},\ }\href {https://doi.org/10.1038/s41586-021-04073-2}
  {\bibfield  {journal} {\bibinfo  {journal} {Nature}\ }\textbf {\bibinfo
  {volume} {601}},\ \bibinfo {pages} {35} (\bibinfo {year} {2022})}\BibitemShut
  {NoStop}%
\bibitem [{\citenamefont {Si}\ \emph {et~al.}(2016)\citenamefont {Si},
  \citenamefont {Yu},\ and\ \citenamefont {Abrahams}}]{2016Si_Abrahams}%
  \BibitemOpen
  \bibfield  {author} {\bibinfo {author} {\bibfnamefont {Q.}~\bibnamefont
  {Si}}, \bibinfo {author} {\bibfnamefont {R.}~\bibnamefont {Yu}},\ and\
  \bibinfo {author} {\bibfnamefont {E.}~\bibnamefont {Abrahams}},\ }\bibfield
  {title} {\bibinfo {title} {High-temperature superconductivity in iron
  pnictides and chalcogenides},\ }\bibfield  {journal} {\bibinfo  {journal}
  {Nature Reviews Materials}\ }\textbf {\bibinfo {volume} {1}},\ \href
  {https://doi.org/10.1038/natrevmats.2016.17} {10.1038/natrevmats.2016.17}
  (\bibinfo {year} {2016})\BibitemShut {NoStop}%
\bibitem [{\citenamefont {Zhang}\ and\ \citenamefont
  {Zhai}(2017)}]{2017Zhang_Zhai}%
  \BibitemOpen
  \bibfield  {author} {\bibinfo {author} {\bibfnamefont {P.}~\bibnamefont
  {Zhang}}\ and\ \bibinfo {author} {\bibfnamefont {H.-f.}\ \bibnamefont
  {Zhai}},\ }\bibfield  {title} {\bibinfo {title} {Superconductivity in
  122-type pnictides without iron},\ }\bibfield  {journal} {\bibinfo  {journal}
  {Condensed Matter}\ }\textbf {\bibinfo {volume} {2}},\ \href
  {https://doi.org/10.3390/condmat2030028} {10.3390/condmat2030028} (\bibinfo
  {year} {2017})\BibitemShut {NoStop}%
\bibitem [{\citenamefont {Shatruk}(2019)}]{2019Shatruk}%
  \BibitemOpen
  \bibfield  {author} {\bibinfo {author} {\bibfnamefont {M.}~\bibnamefont
  {Shatruk}},\ }\bibfield  {title} {\bibinfo {title} {Thcr$_2$si$_2$ structure
  type: The "perovskite" of intermetallics},\ }\href
  {https://doi.org/https://doi.org/10.1016/j.jssc.2019.02.012} {\bibfield
  {journal} {\bibinfo  {journal} {Journal of Solid State Chemistry}\ }\textbf
  {\bibinfo {volume} {272}},\ \bibinfo {pages} {198} (\bibinfo {year}
  {2019})}\BibitemShut {NoStop}%
\bibitem [{\citenamefont {Ban}\ and\ \citenamefont {Sikirica}(1965)}]{1965BAN}%
  \BibitemOpen
  \bibfield  {author} {\bibinfo {author} {\bibfnamefont {Z.}~\bibnamefont
  {Ban}}\ and\ \bibinfo {author} {\bibfnamefont {M.}~\bibnamefont {Sikirica}},\
  }\bibfield  {title} {\bibinfo {title} {{The crystal structure of ternary
  silicides ThM${\sb 2}$Si${\sb 2}$(M = Cr, Mn, Fe, Co, Ni and Cu)}},\
  }\href@noop {} {\bibfield  {journal} {\bibinfo  {journal} {Acta
  Crystallogr.}\ }\textbf {\bibinfo {volume} {18}},\ \bibinfo {pages} {594}
  (\bibinfo {year} {1965})}\BibitemShut {NoStop}%
\bibitem [{\citenamefont {Lu}\ \emph {et~al.}(2016)\citenamefont {Lu},
  \citenamefont {Liu}, \citenamefont {Naumov}, \citenamefont {Meng},
  \citenamefont {Li}, \citenamefont {Tse}, \citenamefont {Yang},\ and\
  \citenamefont {Hemley}}]{2016LU_Hemley}%
  \BibitemOpen
  \bibfield  {author} {\bibinfo {author} {\bibfnamefont {S.}~\bibnamefont
  {Lu}}, \bibinfo {author} {\bibfnamefont {H.}~\bibnamefont {Liu}}, \bibinfo
  {author} {\bibfnamefont {I.~I.}\ \bibnamefont {Naumov}}, \bibinfo {author}
  {\bibfnamefont {S.}~\bibnamefont {Meng}}, \bibinfo {author} {\bibfnamefont
  {Y.}~\bibnamefont {Li}}, \bibinfo {author} {\bibfnamefont {J.~S.}\
  \bibnamefont {Tse}}, \bibinfo {author} {\bibfnamefont {B.}~\bibnamefont
  {Yang}},\ and\ \bibinfo {author} {\bibfnamefont {R.~J.}\ \bibnamefont
  {Hemley}},\ }\bibfield  {title} {\bibinfo {title} {Superconductivity in dense
  carbon-based materials},\ }\href {https://doi.org/10.1103/PhysRevB.93.104509}
  {\bibfield  {journal} {\bibinfo  {journal} {Phys. Rev. B}\ }\textbf {\bibinfo
  {volume} {93}},\ \bibinfo {pages} {104509} (\bibinfo {year}
  {2016})}\BibitemShut {NoStop}%
\bibitem [{\citenamefont {{Sadat Khan}}\ \emph {et~al.}(2022)\citenamefont
  {{Sadat Khan}}, \citenamefont {{Rahman Rano}}, \citenamefont {Syed},
  \citenamefont {Islam},\ and\ \citenamefont {Naqib}}]{2022Khan_Naqib}%
  \BibitemOpen
  \bibfield  {author} {\bibinfo {author} {\bibfnamefont {N.}~\bibnamefont
  {{Sadat Khan}}}, \bibinfo {author} {\bibfnamefont {B.}~\bibnamefont {{Rahman
  Rano}}}, \bibinfo {author} {\bibfnamefont {I.~M.}\ \bibnamefont {Syed}},
  \bibinfo {author} {\bibfnamefont {R.}~\bibnamefont {Islam}},\ and\ \bibinfo
  {author} {\bibfnamefont {S.}~\bibnamefont {Naqib}},\ }\bibfield  {title}
  {\bibinfo {title} {First-principles prediction of pressure dependent
  mechanical, electronic, optical, and superconducting state properties of
  nac6: A potential high-tc superconductor},\ }\href
  {https://doi.org/https://doi.org/10.1016/j.rinp.2022.105182} {\bibfield
  {journal} {\bibinfo  {journal} {Results in Physics}\ }\textbf {\bibinfo
  {volume} {33}},\ \bibinfo {pages} {105182} (\bibinfo {year}
  {2022})}\BibitemShut {NoStop}%
\bibitem [{\citenamefont {Geng}\ \emph {et~al.}(2023)\citenamefont {Geng},
  \citenamefont {Hilleke}, \citenamefont {Zhu}, \citenamefont {Wang},
  \citenamefont {Strobel},\ and\ \citenamefont {Zurek}}]{2023Nisha_Zurek}%
  \BibitemOpen
  \bibfield  {author} {\bibinfo {author} {\bibfnamefont {N.}~\bibnamefont
  {Geng}}, \bibinfo {author} {\bibfnamefont {K.~P.}\ \bibnamefont {Hilleke}},
  \bibinfo {author} {\bibfnamefont {L.}~\bibnamefont {Zhu}}, \bibinfo {author}
  {\bibfnamefont {X.}~\bibnamefont {Wang}}, \bibinfo {author} {\bibfnamefont
  {T.~A.}\ \bibnamefont {Strobel}},\ and\ \bibinfo {author} {\bibfnamefont
  {E.}~\bibnamefont {Zurek}},\ }\bibfield  {title} {\bibinfo {title}
  {Conventional high-temperature superconductivity in metallic, covalently
  bonded, binary-guest c–b clathrates},\ }\href
  {https://doi.org/10.1021/jacs.2c10089} {\bibfield  {journal} {\bibinfo
  {journal} {Journal of the American Chemical Society}\ }\textbf {\bibinfo
  {volume} {145}},\ \bibinfo {pages} {1696} (\bibinfo {year} {2023})},\
  \bibinfo {note} {pMID: 36622785},\ \Eprint
  {https://arxiv.org/abs/https://doi.org/10.1021/jacs.2c10089}
  {https://doi.org/10.1021/jacs.2c10089} \BibitemShut {NoStop}%
\bibitem [{\citenamefont {Dolui}\ \emph {et~al.}(2024)\citenamefont {Dolui},
  \citenamefont {Conway}, \citenamefont {Heil}, \citenamefont {Strobel},
  \citenamefont {Prasankumar},\ and\ \citenamefont
  {Pickard}}]{2024Dolui_Pickard}%
  \BibitemOpen
  \bibfield  {author} {\bibinfo {author} {\bibfnamefont {K.}~\bibnamefont
  {Dolui}}, \bibinfo {author} {\bibfnamefont {L.~J.}\ \bibnamefont {Conway}},
  \bibinfo {author} {\bibfnamefont {C.}~\bibnamefont {Heil}}, \bibinfo {author}
  {\bibfnamefont {T.~A.}\ \bibnamefont {Strobel}}, \bibinfo {author}
  {\bibfnamefont {R.~P.}\ \bibnamefont {Prasankumar}},\ and\ \bibinfo {author}
  {\bibfnamefont {C.~J.}\ \bibnamefont {Pickard}},\ }\bibfield  {title}
  {\bibinfo {title} {Feasible route to high-temperature ambient-pressure
  hydride superconductivity},\ }\href
  {https://doi.org/10.1103/PhysRevLett.132.166001} {\bibfield  {journal}
  {\bibinfo  {journal} {Phys. Rev. Lett.}\ }\textbf {\bibinfo {volume} {132}},\
  \bibinfo {pages} {166001} (\bibinfo {year} {2024})}\BibitemShut {NoStop}%
\bibitem [{\citenamefont {Bergerhoff}\ \emph {et~al.}(1983)\citenamefont
  {Bergerhoff}, \citenamefont {Hundt}, \citenamefont {Sievers},\ and\
  \citenamefont {Brown}}]{1983BER}%
  \BibitemOpen
  \bibfield  {author} {\bibinfo {author} {\bibfnamefont {G.}~\bibnamefont
  {Bergerhoff}}, \bibinfo {author} {\bibfnamefont {R.}~\bibnamefont {Hundt}},
  \bibinfo {author} {\bibfnamefont {R.}~\bibnamefont {Sievers}},\ and\ \bibinfo
  {author} {\bibfnamefont {I.}~\bibnamefont {Brown}},\ }\bibfield  {title}
  {\bibinfo {title} {The inorganic crystal structure data base},\ }\href@noop
  {} {\bibfield  {journal} {\bibinfo  {journal} {J. Chem. Inf. Comput. Sci.}\
  }\textbf {\bibinfo {volume} {23}},\ \bibinfo {pages} {66} (\bibinfo {year}
  {1983})}\BibitemShut {NoStop}%
\bibitem [{\citenamefont {McMillan}(1968)}]{1968MCM}%
  \BibitemOpen
  \bibfield  {author} {\bibinfo {author} {\bibfnamefont {W.~L.}\ \bibnamefont
  {McMillan}},\ }\bibfield  {title} {\bibinfo {title} {Transition temperature
  of strong-coupled superconductors},\ }\href@noop {} {\bibfield  {journal}
  {\bibinfo  {journal} {Phys. Rev.}\ }\textbf {\bibinfo {volume} {167}},\
  \bibinfo {pages} {331} (\bibinfo {year} {1968})}\BibitemShut {NoStop}%
\bibitem [{\citenamefont {Giannozzi}\ \emph {et~al.}(2009)\citenamefont
  {Giannozzi}, \citenamefont {Baroni}, \citenamefont {Bonini}, \citenamefont
  {Calandra}, \citenamefont {Car}, \citenamefont {Cavazzoni}, \citenamefont
  {Ceresoli}, \citenamefont {Chiarotti}, \citenamefont {Cococcioni},
  \citenamefont {Dabo}, \citenamefont {Corso}, \citenamefont {de~Gironcoli},
  \citenamefont {Fabris}, \citenamefont {Fratesi}, \citenamefont {Gebauer},
  \citenamefont {Gerstmann}, \citenamefont {Gougoussis}, \citenamefont
  {Kokalj}, \citenamefont {Lazzeri}, \citenamefont {Martin-Samos},
  \citenamefont {Marzari}, \citenamefont {Mauri}, \citenamefont {Mazzarello},
  \citenamefont {Paolini}, \citenamefont {Pasquarello}, \citenamefont
  {Paulatto}, \citenamefont {Sbraccia}, \citenamefont {Scandolo}, \citenamefont
  {Sclauzero}, \citenamefont {Seitsonen}, \citenamefont {Smogunov},
  \citenamefont {Umari},\ and\ \citenamefont {Wentzcovitch}}]{2009GIA}%
  \BibitemOpen
  \bibfield  {author} {\bibinfo {author} {\bibfnamefont {P.}~\bibnamefont
  {Giannozzi}}, \bibinfo {author} {\bibfnamefont {S.}~\bibnamefont {Baroni}},
  \bibinfo {author} {\bibfnamefont {N.}~\bibnamefont {Bonini}}, \bibinfo
  {author} {\bibfnamefont {M.}~\bibnamefont {Calandra}}, \bibinfo {author}
  {\bibfnamefont {R.}~\bibnamefont {Car}}, \bibinfo {author} {\bibfnamefont
  {C.}~\bibnamefont {Cavazzoni}}, \bibinfo {author} {\bibfnamefont
  {D.}~\bibnamefont {Ceresoli}}, \bibinfo {author} {\bibfnamefont {G.~L.}\
  \bibnamefont {Chiarotti}}, \bibinfo {author} {\bibfnamefont {M.}~\bibnamefont
  {Cococcioni}}, \bibinfo {author} {\bibfnamefont {I.}~\bibnamefont {Dabo}},
  \bibinfo {author} {\bibfnamefont {A.~D.}\ \bibnamefont {Corso}}, \bibinfo
  {author} {\bibfnamefont {S.}~\bibnamefont {de~Gironcoli}}, \bibinfo {author}
  {\bibfnamefont {S.}~\bibnamefont {Fabris}}, \bibinfo {author} {\bibfnamefont
  {G.}~\bibnamefont {Fratesi}}, \bibinfo {author} {\bibfnamefont
  {R.}~\bibnamefont {Gebauer}}, \bibinfo {author} {\bibfnamefont
  {U.}~\bibnamefont {Gerstmann}}, \bibinfo {author} {\bibfnamefont
  {C.}~\bibnamefont {Gougoussis}}, \bibinfo {author} {\bibfnamefont
  {A.}~\bibnamefont {Kokalj}}, \bibinfo {author} {\bibfnamefont
  {M.}~\bibnamefont {Lazzeri}}, \bibinfo {author} {\bibfnamefont
  {L.}~\bibnamefont {Martin-Samos}}, \bibinfo {author} {\bibfnamefont
  {N.}~\bibnamefont {Marzari}}, \bibinfo {author} {\bibfnamefont
  {F.}~\bibnamefont {Mauri}}, \bibinfo {author} {\bibfnamefont
  {R.}~\bibnamefont {Mazzarello}}, \bibinfo {author} {\bibfnamefont
  {S.}~\bibnamefont {Paolini}}, \bibinfo {author} {\bibfnamefont
  {A.}~\bibnamefont {Pasquarello}}, \bibinfo {author} {\bibfnamefont
  {L.}~\bibnamefont {Paulatto}}, \bibinfo {author} {\bibfnamefont
  {C.}~\bibnamefont {Sbraccia}}, \bibinfo {author} {\bibfnamefont
  {S.}~\bibnamefont {Scandolo}}, \bibinfo {author} {\bibfnamefont
  {G.}~\bibnamefont {Sclauzero}}, \bibinfo {author} {\bibfnamefont {A.~P.}\
  \bibnamefont {Seitsonen}}, \bibinfo {author} {\bibfnamefont {A.}~\bibnamefont
  {Smogunov}}, \bibinfo {author} {\bibfnamefont {P.}~\bibnamefont {Umari}},\
  and\ \bibinfo {author} {\bibfnamefont {R.~M.}\ \bibnamefont {Wentzcovitch}},\
  }\bibfield  {title} {\bibinfo {title} {{QUANTUM} {ESPRESSO}: a modular and
  open-source software project for quantum simulations of materials},\
  }\href@noop {} {\bibfield  {journal} {\bibinfo  {journal} {J. Phys.: Condens.
  Matter}\ }\textbf {\bibinfo {volume} {21}},\ \bibinfo {pages} {395502}
  (\bibinfo {year} {2009})}\BibitemShut {NoStop}%
\bibitem [{\citenamefont {Kresse}\ and\ \citenamefont
  {Joubert}(1999)}]{1999GKR}%
  \BibitemOpen
  \bibfield  {author} {\bibinfo {author} {\bibfnamefont {G.}~\bibnamefont
  {Kresse}}\ and\ \bibinfo {author} {\bibfnamefont {D.}~\bibnamefont
  {Joubert}},\ }\bibfield  {title} {\bibinfo {title} {From ultrasoft
  pseudopotentials to the projector augmented-wave method},\ }\href@noop {}
  {\bibfield  {journal} {\bibinfo  {journal} {Phys. Rev. B}\ }\textbf {\bibinfo
  {volume} {59}},\ \bibinfo {pages} {1758} (\bibinfo {year}
  {1999})}\BibitemShut {NoStop}%
\bibitem [{\citenamefont {Kresse}\ and\ \citenamefont
  {Hafner}(1994)}]{1994GKR}%
  \BibitemOpen
  \bibfield  {author} {\bibinfo {author} {\bibfnamefont {G.}~\bibnamefont
  {Kresse}}\ and\ \bibinfo {author} {\bibfnamefont {J.}~\bibnamefont
  {Hafner}},\ }\bibfield  {title} {\bibinfo {title} {Ab initio
  molecular-dynamics simulation of the liquid-metal-amorphous-semiconductor
  transition in germanium},\ }\href@noop {} {\bibfield  {journal} {\bibinfo
  {journal} {Phys. Rev. B}\ }\textbf {\bibinfo {volume} {49}},\ \bibinfo
  {pages} {14251} (\bibinfo {year} {1994})}\BibitemShut {NoStop}%
\bibitem [{\citenamefont {Kresse}\ and\ \citenamefont
  {Furthm{\"u}ller}(1996)}]{1996GKR}%
  \BibitemOpen
  \bibfield  {author} {\bibinfo {author} {\bibfnamefont {G.}~\bibnamefont
  {Kresse}}\ and\ \bibinfo {author} {\bibfnamefont {J.}~\bibnamefont
  {Furthm{\"u}ller}},\ }\bibfield  {title} {\bibinfo {title} {Efficiency of
  ab-initio total energy calculations for metals and semiconductors using a
  plane-wave basis set},\ }\href@noop {} {\bibfield  {journal} {\bibinfo
  {journal} {Comput. Mater. Sci.}\ }\textbf {\bibinfo {volume} {6}},\ \bibinfo
  {pages} {15} (\bibinfo {year} {1996})}\BibitemShut {NoStop}%
\bibitem [{\citenamefont {Vanderbilt}(1990)}]{1990DV}%
  \BibitemOpen
  \bibfield  {author} {\bibinfo {author} {\bibfnamefont {D.}~\bibnamefont
  {Vanderbilt}},\ }\bibfield  {title} {\bibinfo {title} {Soft self-consistent
  pseudopotentials in a generalized eigenvalue formalism},\ }\href
  {https://doi.org/10.1103/PhysRevB.41.7892} {\bibfield  {journal} {\bibinfo
  {journal} {Phys. Rev. B}\ }\textbf {\bibinfo {volume} {41}},\ \bibinfo
  {pages} {7892} (\bibinfo {year} {1990})}\BibitemShut {NoStop}%
\bibitem [{\citenamefont {{Dal Corso}}(2014)}]{2014DAL}%
  \BibitemOpen
  \bibfield  {author} {\bibinfo {author} {\bibfnamefont {A.}~\bibnamefont {{Dal
  Corso}}},\ }\bibfield  {title} {\bibinfo {title} {{Pseudopotentials periodic
  table: From H to Pu}},\ }\href@noop {} {\bibfield  {journal} {\bibinfo
  {journal} {Comput. Mater. Sci.}\ }\textbf {\bibinfo {volume} {95}},\ \bibinfo
  {pages} {337} (\bibinfo {year} {2014})}\BibitemShut {NoStop}%
\bibitem [{\citenamefont {Marzari}\ \emph {et~al.}(1999)\citenamefont
  {Marzari}, \citenamefont {Vanderbilt}, \citenamefont {De~Vita},\ and\
  \citenamefont {Payne}}]{1999MAR}%
  \BibitemOpen
  \bibfield  {author} {\bibinfo {author} {\bibfnamefont {N.}~\bibnamefont
  {Marzari}}, \bibinfo {author} {\bibfnamefont {D.}~\bibnamefont {Vanderbilt}},
  \bibinfo {author} {\bibfnamefont {A.}~\bibnamefont {De~Vita}},\ and\ \bibinfo
  {author} {\bibfnamefont {M.~C.}\ \bibnamefont {Payne}},\ }\bibfield  {title}
  {\bibinfo {title} {Thermal contraction and disordering of the al(110)
  surface},\ }\href@noop {} {\bibfield  {journal} {\bibinfo  {journal} {Phys.
  Rev. Lett.}\ }\textbf {\bibinfo {volume} {82}},\ \bibinfo {pages} {3296}
  (\bibinfo {year} {1999})}\BibitemShut {NoStop}%
\bibitem [{\citenamefont {Dronskowski}\ and\ \citenamefont
  {Bloechl}(1993)}]{1993Dronskowski_Bloechl}%
  \BibitemOpen
  \bibfield  {author} {\bibinfo {author} {\bibfnamefont {R.}~\bibnamefont
  {Dronskowski}}\ and\ \bibinfo {author} {\bibfnamefont {P.~E.}\ \bibnamefont
  {Bloechl}},\ }\bibfield  {title} {\bibinfo {title} {Crystal orbital hamilton
  populations (cohp): energy-resolved visualization of chemical bonding in
  solids based on density-functional calculations},\ }\href
  {https://doi.org/10.1021/j100135a014} {\bibfield  {journal} {\bibinfo
  {journal} {The Journal of Physical Chemistry}\ }\textbf {\bibinfo {volume}
  {97}},\ \bibinfo {pages} {8617} (\bibinfo {year} {1993})}\BibitemShut
  {NoStop}%
\bibitem [{\citenamefont {Deringer}\ \emph {et~al.}(2011)\citenamefont
  {Deringer}, \citenamefont {Tchougr{\'e}eff},\ and\ \citenamefont
  {Dronskowski}}]{2011Deringer_Dronskowski}%
  \BibitemOpen
  \bibfield  {author} {\bibinfo {author} {\bibfnamefont {V.~L.}\ \bibnamefont
  {Deringer}}, \bibinfo {author} {\bibfnamefont {A.~L.}\ \bibnamefont
  {Tchougr{\'e}eff}},\ and\ \bibinfo {author} {\bibfnamefont {R.}~\bibnamefont
  {Dronskowski}},\ }\bibfield  {title} {\bibinfo {title} {Crystal orbital
  hamilton population (cohp) analysis as projected from plane-wave basis
  sets},\ }\href {https://doi.org/10.1021/jp202489s} {\bibfield  {journal}
  {\bibinfo  {journal} {The Journal of Physical Chemistry A}\ }\textbf
  {\bibinfo {volume} {115}},\ \bibinfo {pages} {5461} (\bibinfo {year}
  {2011})},\ \bibinfo {note} {pMID: 21548594},\ \Eprint
  {https://arxiv.org/abs/https://doi.org/10.1021/jp202489s}
  {https://doi.org/10.1021/jp202489s} \BibitemShut {NoStop}%
\bibitem [{\citenamefont {Maintz}\ \emph {et~al.}(2013)\citenamefont {Maintz},
  \citenamefont {Deringer}, \citenamefont {Tchougréeff},\ and\ \citenamefont
  {Dronskowski}}]{2013Maintz_Dronskowski}%
  \BibitemOpen
  \bibfield  {author} {\bibinfo {author} {\bibfnamefont {S.}~\bibnamefont
  {Maintz}}, \bibinfo {author} {\bibfnamefont {V.~L.}\ \bibnamefont
  {Deringer}}, \bibinfo {author} {\bibfnamefont {A.~L.}\ \bibnamefont
  {Tchougréeff}},\ and\ \bibinfo {author} {\bibfnamefont {R.}~\bibnamefont
  {Dronskowski}},\ }\bibfield  {title} {\bibinfo {title} {Analytic projection
  from plane-wave and paw wavefunctions and application to chemical-bonding
  analysis in solids},\ }\href
  {https://doi.org/https://doi.org/10.1002/jcc.23424} {\bibfield  {journal}
  {\bibinfo  {journal} {Journal of Computational Chemistry}\ }\textbf {\bibinfo
  {volume} {34}},\ \bibinfo {pages} {2557} (\bibinfo {year} {2013})},\ \Eprint
  {https://arxiv.org/abs/https://onlinelibrary.wiley.com/doi/pdf/10.1002/jcc.23424}
  {https://onlinelibrary.wiley.com/doi/pdf/10.1002/jcc.23424} \BibitemShut
  {NoStop}%
\bibitem [{\citenamefont {Maintz}\ \emph {et~al.}(2016)\citenamefont {Maintz},
  \citenamefont {Deringer}, \citenamefont {Tchougréeff},\ and\ \citenamefont
  {Dronskowski}}]{2016Maintz_Dronskowski}%
  \BibitemOpen
  \bibfield  {author} {\bibinfo {author} {\bibfnamefont {S.}~\bibnamefont
  {Maintz}}, \bibinfo {author} {\bibfnamefont {V.~L.}\ \bibnamefont
  {Deringer}}, \bibinfo {author} {\bibfnamefont {A.~L.}\ \bibnamefont
  {Tchougréeff}},\ and\ \bibinfo {author} {\bibfnamefont {R.}~\bibnamefont
  {Dronskowski}},\ }\bibfield  {title} {\bibinfo {title} {Lobster: A tool to
  extract chemical bonding from plane-wave based dft},\ }\href
  {https://doi.org/https://doi.org/10.1002/jcc.24300} {\bibfield  {journal}
  {\bibinfo  {journal} {Journal of Computational Chemistry}\ }\textbf {\bibinfo
  {volume} {37}},\ \bibinfo {pages} {1030} (\bibinfo {year} {2016})},\ \Eprint
  {https://arxiv.org/abs/https://onlinelibrary.wiley.com/doi/pdf/10.1002/jcc.24300}
  {https://onlinelibrary.wiley.com/doi/pdf/10.1002/jcc.24300} \BibitemShut
  {NoStop}%
\bibitem [{\citenamefont {Nelson}\ \emph {et~al.}(2020)\citenamefont {Nelson},
  \citenamefont {Ertural}, \citenamefont {George}, \citenamefont {Deringer},
  \citenamefont {Hautier},\ and\ \citenamefont
  {Dronskowski}}]{2020Nelson_Dronskowski}%
  \BibitemOpen
  \bibfield  {author} {\bibinfo {author} {\bibfnamefont {R.}~\bibnamefont
  {Nelson}}, \bibinfo {author} {\bibfnamefont {C.}~\bibnamefont {Ertural}},
  \bibinfo {author} {\bibfnamefont {J.}~\bibnamefont {George}}, \bibinfo
  {author} {\bibfnamefont {V.~L.}\ \bibnamefont {Deringer}}, \bibinfo {author}
  {\bibfnamefont {G.}~\bibnamefont {Hautier}},\ and\ \bibinfo {author}
  {\bibfnamefont {R.}~\bibnamefont {Dronskowski}},\ }\bibfield  {title}
  {\bibinfo {title} {Lobster: Local orbital projections, atomic charges, and
  chemical-bonding analysis from projector-augmented-wave-based
  density-functional theory},\ }\href
  {https://doi.org/https://doi.org/10.1002/jcc.26353} {\bibfield  {journal}
  {\bibinfo  {journal} {Journal of Computational Chemistry}\ }\textbf {\bibinfo
  {volume} {41}},\ \bibinfo {pages} {1931} (\bibinfo {year} {2020})},\ \Eprint
  {https://arxiv.org/abs/https://onlinelibrary.wiley.com/doi/pdf/10.1002/jcc.26353}
  {https://onlinelibrary.wiley.com/doi/pdf/10.1002/jcc.26353} \BibitemShut
  {NoStop}%
\bibitem [{\citenamefont {Kawamura}(2019)}]{2019Kawamura}%
  \BibitemOpen
  \bibfield  {author} {\bibinfo {author} {\bibfnamefont {M.}~\bibnamefont
  {Kawamura}},\ }\bibfield  {title} {\bibinfo {title} {Fermisurfer:
  Fermi-surface viewer providing multiple representation schemes},\ }\href
  {https://doi.org/https://doi.org/10.1016/j.cpc.2019.01.017} {\bibfield
  {journal} {\bibinfo  {journal} {Computer Physics Communications}\ }\textbf
  {\bibinfo {volume} {239}},\ \bibinfo {pages} {197} (\bibinfo {year}
  {2019})}\BibitemShut {NoStop}%
\bibitem [{\citenamefont {Wierzbowska}\ \emph {et~al.}(2006)\citenamefont
  {Wierzbowska}, \citenamefont {de~Gironcoli},\ and\ \citenamefont
  {Giannozzi}}]{2006WIERZBOWSKA_GIANNOZZI}%
  \BibitemOpen
  \bibfield  {author} {\bibinfo {author} {\bibfnamefont {M.}~\bibnamefont
  {Wierzbowska}}, \bibinfo {author} {\bibfnamefont {S.}~\bibnamefont
  {de~Gironcoli}},\ and\ \bibinfo {author} {\bibfnamefont {P.}~\bibnamefont
  {Giannozzi}},\ }\href {https://arxiv.org/abs/cond-mat/0504077} {\bibinfo
  {title} {Origins of low- and high-pressure discontinuities of $t_{c}$ in
  niobium}} (\bibinfo {year} {2006}),\ \Eprint
  {https://arxiv.org/abs/cond-mat/0504077} {arXiv:cond-mat/0504077
  [cond-mat.supr-con]} \BibitemShut {NoStop}%
\bibitem [{\citenamefont {Venturini}\ \emph {et~al.}(1989)\citenamefont
  {Venturini}, \citenamefont {Malaman},\ and\ \citenamefont
  {Roques}}]{1989VEN}%
  \BibitemOpen
  \bibfield  {author} {\bibinfo {author} {\bibfnamefont {G.}~\bibnamefont
  {Venturini}}, \bibinfo {author} {\bibfnamefont {B.}~\bibnamefont {Malaman}},\
  and\ \bibinfo {author} {\bibfnamefont {B.}~\bibnamefont {Roques}},\
  }\bibfield  {title} {\bibinfo {title} {Lapt$_2$ge$_2$, a monoclinic variant
  of the tetragonal cabe$_2$ge$_2$-type structure},\ }\href@noop {} {\bibfield
  {journal} {\bibinfo  {journal} {J. Less-Common Met.}\ }\textbf {\bibinfo
  {volume} {146}},\ \bibinfo {pages} {271} (\bibinfo {year}
  {1989})}\BibitemShut {NoStop}%
\bibitem [{\citenamefont {Marazza}\ \emph {et~al.}(1977)\citenamefont
  {Marazza}, \citenamefont {Ferro}, \citenamefont {Rambaldi},\ and\
  \citenamefont {Zanicchi}}]{1977MAR}%
  \BibitemOpen
  \bibfield  {author} {\bibinfo {author} {\bibfnamefont {R.}~\bibnamefont
  {Marazza}}, \bibinfo {author} {\bibfnamefont {R.}~\bibnamefont {Ferro}},
  \bibinfo {author} {\bibfnamefont {G.}~\bibnamefont {Rambaldi}},\ and\
  \bibinfo {author} {\bibfnamefont {G.}~\bibnamefont {Zanicchi}},\ }\bibfield
  {title} {\bibinfo {title} {Some phases in ternary alloys of thorium and
  uranium with the al4ba-thcu2si2-type structure},\ }\href@noop {} {\bibfield
  {journal} {\bibinfo  {journal} {J. Less-Common Met.}\ }\textbf {\bibinfo
  {volume} {53}},\ \bibinfo {pages} {193} (\bibinfo {year} {1977})}\BibitemShut
  {NoStop}%
\bibitem [{\citenamefont {Shelton}\ \emph {et~al.}(1984)\citenamefont
  {Shelton}, \citenamefont {Braun},\ and\ \citenamefont {Musick}}]{1984SHE}%
  \BibitemOpen
  \bibfield  {author} {\bibinfo {author} {\bibfnamefont {R.}~\bibnamefont
  {Shelton}}, \bibinfo {author} {\bibfnamefont {H.}~\bibnamefont {Braun}},\
  and\ \bibinfo {author} {\bibfnamefont {E.}~\bibnamefont {Musick}},\
  }\bibfield  {title} {\bibinfo {title} {Superconductivity and relative phase
  stability in 1:2:2 ternary transition metal silicides and germanides},\
  }\href@noop {} {\bibfield  {journal} {\bibinfo  {journal} {Solid State
  Commun.}\ }\textbf {\bibinfo {volume} {52}},\ \bibinfo {pages} {797}
  (\bibinfo {year} {1984})}\BibitemShut {NoStop}%
\bibitem [{\citenamefont {Hase}\ and\ \citenamefont
  {Yanagisawa}(2013)}]{2013HAS}%
  \BibitemOpen
  \bibfield  {author} {\bibinfo {author} {\bibfnamefont {I.}~\bibnamefont
  {Hase}}\ and\ \bibinfo {author} {\bibfnamefont {T.}~\bibnamefont
  {Yanagisawa}},\ }\bibfield  {title} {\bibinfo {title} {Electronic structure
  of lapt$_2$si$_2$},\ }\href@noop {} {\bibfield  {journal} {\bibinfo
  {journal} {Phys. C (Amsterdam, Neth.)}\ }\textbf {\bibinfo {volume} {484}},\
  \bibinfo {pages} {59} (\bibinfo {year} {2013})}\BibitemShut {NoStop}%
\bibitem [{\citenamefont {Zada}\ \emph {et~al.}(2020)\citenamefont {Zada},
  \citenamefont {Ullah}, \citenamefont {Bibi}, \citenamefont {Zada},\ and\
  \citenamefont {Mahmood}}]{2020ZAD}%
  \BibitemOpen
  \bibfield  {author} {\bibinfo {author} {\bibfnamefont {Z.}~\bibnamefont
  {Zada}}, \bibinfo {author} {\bibfnamefont {H.}~\bibnamefont {Ullah}},
  \bibinfo {author} {\bibfnamefont {R.}~\bibnamefont {Bibi}}, \bibinfo {author}
  {\bibfnamefont {S.}~\bibnamefont {Zada}},\ and\ \bibinfo {author}
  {\bibfnamefont {A.}~\bibnamefont {Mahmood}},\ }\bibfield  {title} {\bibinfo
  {title} {Electronic band profiles and magneto-electronic properties of
  ternary xcu$_2$p$_2$ (x= ca, sr) compounds: insight from ab initio
  calculations},\ }\href@noop {} {\bibfield  {journal} {\bibinfo  {journal}
  {Zeitschrift f{\"u}r Naturforschung A}\ }\textbf {\bibinfo {volume} {75}},\
  \bibinfo {pages} {543} (\bibinfo {year} {2020})}\BibitemShut {NoStop}%
\bibitem [{\citenamefont {Nakano}\ \emph {et~al.}(2016)\citenamefont {Nakano},
  \citenamefont {Hongo},\ and\ \citenamefont {Maezono}}]{2016Nakano_Maezono}%
  \BibitemOpen
  \bibfield  {author} {\bibinfo {author} {\bibfnamefont {K.}~\bibnamefont
  {Nakano}}, \bibinfo {author} {\bibfnamefont {K.}~\bibnamefont {Hongo}},\ and\
  \bibinfo {author} {\bibfnamefont {R.}~\bibnamefont {Maezono}},\ }\bibfield
  {title} {\bibinfo {title} {Phonon dispersions and fermi surfaces nesting
  explaining the variety of charge ordering in titanium-oxypnictides
  superconductors},\ }\href {https://doi.org/10.1038/srep29661} {\bibfield
  {journal} {\bibinfo  {journal} {Scientific Reports}\ }\textbf {\bibinfo
  {volume} {6}},\ \bibinfo {pages} {29661} (\bibinfo {year}
  {2016})}\BibitemShut {NoStop}%
\bibitem [{\citenamefont {Nakano}\ \emph {et~al.}(2017)\citenamefont {Nakano},
  \citenamefont {Hongo},\ and\ \citenamefont {Maezono}}]{2017Nakano_Maezono}%
  \BibitemOpen
  \bibfield  {author} {\bibinfo {author} {\bibfnamefont {K.}~\bibnamefont
  {Nakano}}, \bibinfo {author} {\bibfnamefont {K.}~\bibnamefont {Hongo}},\ and\
  \bibinfo {author} {\bibfnamefont {R.}~\bibnamefont {Maezono}},\ }\bibfield
  {title} {\bibinfo {title} {Investigation into structural phase transitions in
  layered titanium-oxypnictides by a computational phonon analysis},\ }\href
  {https://doi.org/10.1021/acs.inorgchem.7b01709} {\bibfield  {journal}
  {\bibinfo  {journal} {Inorganic Chemistry}\ }\textbf {\bibinfo {volume}
  {56}},\ \bibinfo {pages} {13732} (\bibinfo {year} {2017})},\ \bibinfo {note}
  {pMID: 29094926},\ \Eprint
  {https://arxiv.org/abs/https://doi.org/10.1021/acs.inorgchem.7b01709}
  {https://doi.org/10.1021/acs.inorgchem.7b01709} \BibitemShut {NoStop}%
\bibitem [{\citenamefont {Guo}\ \emph {et~al.}(2016)\citenamefont {Guo},
  \citenamefont {Pan}, \citenamefont {Yu}, \citenamefont {Ruan}, \citenamefont
  {Chen}, \citenamefont {Wang}, \citenamefont {Mu}, \citenamefont {Chen},\ and\
  \citenamefont {Ren}}]{2016QG_ZAR}%
  \BibitemOpen
  \bibfield  {author} {\bibinfo {author} {\bibfnamefont {Q.}~\bibnamefont
  {Guo}}, \bibinfo {author} {\bibfnamefont {B.~J.}\ \bibnamefont {Pan}},
  \bibinfo {author} {\bibfnamefont {J.}~\bibnamefont {Yu}}, \bibinfo {author}
  {\bibfnamefont {B.~B.}\ \bibnamefont {Ruan}}, \bibinfo {author}
  {\bibfnamefont {D.~Y.}\ \bibnamefont {Chen}}, \bibinfo {author}
  {\bibfnamefont {X.~C.}\ \bibnamefont {Wang}}, \bibinfo {author}
  {\bibfnamefont {Q.~G.}\ \bibnamefont {Mu}}, \bibinfo {author} {\bibfnamefont
  {G.~F.}\ \bibnamefont {Chen}},\ and\ \bibinfo {author} {\bibfnamefont
  {Z.~A.}\ \bibnamefont {Ren}},\ }\bibfield  {title} {\bibinfo {title}
  {{Superconductivity at 7.8 K in the ternary LaRu$_2$As$_2$ compound}},\
  }\href {https://doi.org/10.1007/s11434-016-1080-4} {\bibfield  {journal}
  {\bibinfo  {journal} {Science Bulletin}\ }\textbf {\bibinfo {volume} {61}},\
  \bibinfo {pages} {921} (\bibinfo {year} {2016})}\BibitemShut {NoStop}%
\bibitem [{Note1()}]{Note1}%
  \BibitemOpen
  \bibinfo {note} {LaRu$_2$As$_2$ has been suggested to be a multi-band
  superconductor with multiple bands contributing to superconductivity, as
  indicated by DFT calculations \cite {2017Hadi_Islam}.}\BibitemShut {Stop}%
\bibitem [{\citenamefont {Jeitschko}\ \emph {et~al.}(1987)\citenamefont
  {Jeitschko}, \citenamefont {Glaum},\ and\ \citenamefont {Boonk}}]{1987WJ_LB}%
  \BibitemOpen
  \bibfield  {author} {\bibinfo {author} {\bibfnamefont {W.}~\bibnamefont
  {Jeitschko}}, \bibinfo {author} {\bibfnamefont {R.}~\bibnamefont {Glaum}},\
  and\ \bibinfo {author} {\bibfnamefont {L.}~\bibnamefont {Boonk}},\ }\bibfield
   {title} {\bibinfo {title} {{Superconducting LaRu$_2$P$_2$ and other alkaline
  earth and rare earth metal ruthenium and osmium phosphides and arsenides with
  ThCr$_2$Si$_2$ structure}},\ }\href
  {https://doi.org/10.1016/0022-4596(87)90014-4} {\bibfield  {journal}
  {\bibinfo  {journal} {Journal of Solid State Chemistry}\ }\textbf {\bibinfo
  {volume} {69}},\ \bibinfo {pages} {93} (\bibinfo {year} {1987})}\BibitemShut
  {NoStop}%
\bibitem [{\citenamefont {Felner}\ and\ \citenamefont
  {Nowik}(1984)}]{1984IF_IN}%
  \BibitemOpen
  \bibfield  {author} {\bibinfo {author} {\bibfnamefont {I.}~\bibnamefont
  {Felner}}\ and\ \bibinfo {author} {\bibfnamefont {I.}~\bibnamefont {Nowik}},\
  }\bibfield  {title} {\bibinfo {title} {{Itinerant and local magnetism,
  superconductivity and mixed valency phenomena in $RM_2$Si$_2$, ($R$ = rare
  earth, $M$ = Rh, Ru)}},\ }\href
  {https://doi.org/10.1016/0022-3697(84)90149-5} {\bibfield  {journal}
  {\bibinfo  {journal} {Journal of Physics and Chemistry of Solids}\ }\textbf
  {\bibinfo {volume} {45}},\ \bibinfo {pages} {419} (\bibinfo {year}
  {1984})}\BibitemShut {NoStop}%
\bibitem [{\citenamefont {Felner}\ and\ \citenamefont
  {Nowik}(1983)}]{1983IF_IN}%
  \BibitemOpen
  \bibfield  {author} {\bibinfo {author} {\bibfnamefont {I.}~\bibnamefont
  {Felner}}\ and\ \bibinfo {author} {\bibfnamefont {I.}~\bibnamefont {Nowik}},\
  }\bibfield  {title} {\bibinfo {title} {{Local and itinerant magnetism and
  superconductivity in $R$Rh$_2$Si$_2$ ($R$ = rare earth)}},\ }\href
  {https://doi.org/https://doi.org/10.1016/0038-1098(83)90076-5} {\bibfield
  {journal} {\bibinfo  {journal} {Solid State Communications}\ }\textbf
  {\bibinfo {volume} {47}},\ \bibinfo {pages} {831} (\bibinfo {year}
  {1983})}\BibitemShut {NoStop}%
\bibitem [{\citenamefont {Palstra}\ \emph {et~al.}(1986)\citenamefont
  {Palstra}, \citenamefont {Lu}, \citenamefont {Menovsky}, \citenamefont
  {Nieuwenhuys}, \citenamefont {Kes},\ and\ \citenamefont
  {Mydosh}}]{1986TTMP_JAM}%
  \BibitemOpen
  \bibfield  {author} {\bibinfo {author} {\bibfnamefont {T.~T.}\ \bibnamefont
  {Palstra}}, \bibinfo {author} {\bibfnamefont {G.}~\bibnamefont {Lu}},
  \bibinfo {author} {\bibfnamefont {A.~A.}\ \bibnamefont {Menovsky}}, \bibinfo
  {author} {\bibfnamefont {G.~J.}\ \bibnamefont {Nieuwenhuys}}, \bibinfo
  {author} {\bibfnamefont {P.~H.}\ \bibnamefont {Kes}},\ and\ \bibinfo {author}
  {\bibfnamefont {J.~A.}\ \bibnamefont {Mydosh}},\ }\bibfield  {title}
  {\bibinfo {title} {{Superconductivity in the ternary rare-earth (Y, La, and
  Lu) compounds $R$Pd$_2$Si$_2$ and $R$Rh$_2$Si$_2$}},\ }\href
  {https://doi.org/10.1103/PhysRevB.34.4566} {\bibfield  {journal} {\bibinfo
  {journal} {Physical Review B}\ }\textbf {\bibinfo {volume} {34}},\ \bibinfo
  {pages} {4566} (\bibinfo {year} {1986})}\BibitemShut {NoStop}%
\bibitem [{\citenamefont {Chajewski}\ \emph {et~al.}(2018)\citenamefont
  {Chajewski}, \citenamefont {Samsel-Czeka{\l}a}, \citenamefont {Hackemer},
  \citenamefont {Wi{\'s}niewski}, \citenamefont {Pikul},\ and\ \citenamefont
  {Kaczorowski}}]{2018CHA}%
  \BibitemOpen
  \bibfield  {author} {\bibinfo {author} {\bibfnamefont {G.}~\bibnamefont
  {Chajewski}}, \bibinfo {author} {\bibfnamefont {M.}~\bibnamefont
  {Samsel-Czeka{\l}a}}, \bibinfo {author} {\bibfnamefont {A.}~\bibnamefont
  {Hackemer}}, \bibinfo {author} {\bibfnamefont {P.}~\bibnamefont
  {Wi{\'s}niewski}}, \bibinfo {author} {\bibfnamefont {A.}~\bibnamefont
  {Pikul}},\ and\ \bibinfo {author} {\bibfnamefont {D.}~\bibnamefont
  {Kaczorowski}},\ }\bibfield  {title} {\bibinfo {title} {{Superconductivity in
  Y$TE_2$Ge$_2$ compounds ($TE$ = $d$-electron transition metal)}},\
  }\href@noop {} {\bibfield  {journal} {\bibinfo  {journal} {Physica B Condens.
  Matter}\ }\textbf {\bibinfo {volume} {536}},\ \bibinfo {pages} {767}
  (\bibinfo {year} {2018})}\BibitemShut {NoStop}%
\bibitem [{\citenamefont {Hull}\ \emph {et~al.}(1981)\citenamefont {Hull},
  \citenamefont {Wernick}, \citenamefont {Geballe}, \citenamefont {Waszczak},\
  and\ \citenamefont {Bernardini}}]{1981GWH_JEB}%
  \BibitemOpen
  \bibfield  {author} {\bibinfo {author} {\bibfnamefont {G.~W.}\ \bibnamefont
  {Hull}}, \bibinfo {author} {\bibfnamefont {J.~H.}\ \bibnamefont {Wernick}},
  \bibinfo {author} {\bibfnamefont {T.~H.}\ \bibnamefont {Geballe}}, \bibinfo
  {author} {\bibfnamefont {J.~V.}\ \bibnamefont {Waszczak}},\ and\ \bibinfo
  {author} {\bibfnamefont {J.~E.}\ \bibnamefont {Bernardini}},\ }\bibfield
  {title} {\bibinfo {title} {{Superconductivity in the ternary intermetallics
  Yb${\mathrm{Pd}}_{2}$${\mathrm{Ge}}_{2}$,
  La${\mathrm{Pd}}_{2}$${\mathrm{Ge}}_{2}$, and
  La${\mathrm{Pt}}_{2}$${\mathrm{Ge}}_{2}$}},\ }\href
  {https://doi.org/10.1103/PhysRevB.24.6715} {\bibfield  {journal} {\bibinfo
  {journal} {Phys. Rev. B}\ }\textbf {\bibinfo {volume} {24}},\ \bibinfo
  {pages} {6715} (\bibinfo {year} {1981})}\BibitemShut {NoStop}%
\bibitem [{\citenamefont {Hirai}\ \emph {et~al.}(2010)\citenamefont {Hirai},
  \citenamefont {Takayama}, \citenamefont {Hashizume}, \citenamefont
  {Higashinaka}, \citenamefont {Yamamoto}, \citenamefont {Hiroko},\ and\
  \citenamefont {Takagi}}]{2010DH_HT}%
  \BibitemOpen
  \bibfield  {author} {\bibinfo {author} {\bibfnamefont {D.}~\bibnamefont
  {Hirai}}, \bibinfo {author} {\bibfnamefont {T.}~\bibnamefont {Takayama}},
  \bibinfo {author} {\bibfnamefont {D.}~\bibnamefont {Hashizume}}, \bibinfo
  {author} {\bibfnamefont {R.}~\bibnamefont {Higashinaka}}, \bibinfo {author}
  {\bibfnamefont {A.}~\bibnamefont {Yamamoto}}, \bibinfo {author}
  {\bibfnamefont {A.}~\bibnamefont {Hiroko}},\ and\ \bibinfo {author}
  {\bibfnamefont {H.}~\bibnamefont {Takagi}},\ }\bibfield  {title} {\bibinfo
  {title} {{Superconductivity in 4d and 5d transition metal layered pnictides
  BaRh$_2$P$_2$, BaIr$_2$P$_2$ and SrIr$_2$As$_2$}},\ }\href
  {https://doi.org/https://doi.org/10.1016/j.physc.2009.11.059} {\bibfield
  {journal} {\bibinfo  {journal} {Physica C: Superconductivity and its
  Applications}\ }\textbf {\bibinfo {volume} {470}},\ \bibinfo {pages} {S296}
  (\bibinfo {year} {2010})}\BibitemShut {NoStop}%
\bibitem [{\citenamefont {Blawat}\ \emph {et~al.}(2020)\citenamefont {Blawat},
  \citenamefont {Swatek}, \citenamefont {Das}, \citenamefont {Kaczorowski},
  \citenamefont {Jin},\ and\ \citenamefont {Xie}}]{2020JB_WX}%
  \BibitemOpen
  \bibfield  {author} {\bibinfo {author} {\bibfnamefont {J.}~\bibnamefont
  {Blawat}}, \bibinfo {author} {\bibfnamefont {P.~W.}\ \bibnamefont {Swatek}},
  \bibinfo {author} {\bibfnamefont {D.}~\bibnamefont {Das}}, \bibinfo {author}
  {\bibfnamefont {D.}~\bibnamefont {Kaczorowski}}, \bibinfo {author}
  {\bibfnamefont {R.}~\bibnamefont {Jin}},\ and\ \bibinfo {author}
  {\bibfnamefont {W.}~\bibnamefont {Xie}},\ }\bibfield  {title} {\bibinfo
  {title} {{Pd-P antibonding interactions in $A$Pd$_2$P$_2$ ($A$ = Ca and Sr)
  superconductors}},\ }\href
  {https://doi.org/10.1103/PhysRevMaterials.4.014801} {\bibfield  {journal}
  {\bibinfo  {journal} {Physical Review Materials}\ }\textbf {\bibinfo {volume}
  {4}},\ \bibinfo {pages} {14801} (\bibinfo {year} {2020})}\BibitemShut
  {NoStop}%
\bibitem [{\citenamefont {Fujii}\ and\ \citenamefont {Sato}(2009)}]{2009HF_AS}%
  \BibitemOpen
  \bibfield  {author} {\bibinfo {author} {\bibfnamefont {H.}~\bibnamefont
  {Fujii}}\ and\ \bibinfo {author} {\bibfnamefont {A.}~\bibnamefont {Sato}},\
  }\bibfield  {title} {\bibinfo {title} {{Superconductivity in
  SrPd$_2$Ge$_2$}},\ }\href {https://doi.org/10.1103/PhysRevB.79.224522}
  {\bibfield  {journal} {\bibinfo  {journal} {Physical Review B}\ }\textbf
  {\bibinfo {volume} {79}},\ \bibinfo {pages} {1} (\bibinfo {year}
  {2009})}\BibitemShut {NoStop}%
\bibitem [{\citenamefont {Drachuck}\ \emph {et~al.}(2016)\citenamefont
  {Drachuck}, \citenamefont {B{\"{o}}hmer}, \citenamefont {Bud'ko},\ and\
  \citenamefont {Canfield}}]{2016GD_PCC}%
  \BibitemOpen
  \bibfield  {author} {\bibinfo {author} {\bibfnamefont {G.}~\bibnamefont
  {Drachuck}}, \bibinfo {author} {\bibfnamefont {A.~E.}\ \bibnamefont
  {B{\"{o}}hmer}}, \bibinfo {author} {\bibfnamefont {S.~L.}\ \bibnamefont
  {Bud'ko}},\ and\ \bibinfo {author} {\bibfnamefont {P.~C.}\ \bibnamefont
  {Canfield}},\ }\bibfield  {title} {\bibinfo {title} {{Magnetization and
  transport properties of single crystalline $R$Pd$_2$P$_2$ ($R$=Y, La–Nd,
  Sm–Ho, Yb)}},\ }\href {https://doi.org/10.1016/j.jmmm.2016.05.089}
  {\bibfield  {journal} {\bibinfo  {journal} {Journal of Magnetism and Magnetic
  Materials}\ }\textbf {\bibinfo {volume} {417}},\ \bibinfo {pages} {420}
  (\bibinfo {year} {2016})}\BibitemShut {NoStop}%
\bibitem [{\citenamefont {Anand}\ \emph {et~al.}(2013)\citenamefont {Anand},
  \citenamefont {Kim}, \citenamefont {Tanatar}, \citenamefont {Prozorov},\ and\
  \citenamefont {Johnston}}]{2013VKA_DCJ}%
  \BibitemOpen
  \bibfield  {author} {\bibinfo {author} {\bibfnamefont {V.~K.}\ \bibnamefont
  {Anand}}, \bibinfo {author} {\bibfnamefont {H.}~\bibnamefont {Kim}}, \bibinfo
  {author} {\bibfnamefont {M.~A.}\ \bibnamefont {Tanatar}}, \bibinfo {author}
  {\bibfnamefont {R.}~\bibnamefont {Prozorov}},\ and\ \bibinfo {author}
  {\bibfnamefont {D.~C.}\ \bibnamefont {Johnston}},\ }\bibfield  {title}
  {\bibinfo {title} {{Superconducting and normal-state properties of
  $A$Pd$_2$As$_2$ ($A$ = Ca, Sr, Ba) single crystals}},\ }\href
  {https://doi.org/10.1103/PhysRevB.87.224510} {\bibfield  {journal} {\bibinfo
  {journal} {Physical Review B}\ }\textbf {\bibinfo {volume} {87}},\ \bibinfo
  {pages} {1} (\bibinfo {year} {2013})}\BibitemShut {NoStop}%
\bibitem [{\citenamefont {Wang}\ \emph {et~al.}(2017)\citenamefont {Wang},
  \citenamefont {Ruan}, \citenamefont {Yu}, \citenamefont {Pan}, \citenamefont
  {Mu}, \citenamefont {Liu}, \citenamefont {Chen},\ and\ \citenamefont
  {Ren}}]{2017WANG_REN}%
  \BibitemOpen
  \bibfield  {author} {\bibinfo {author} {\bibfnamefont {X.-C.}\ \bibnamefont
  {Wang}}, \bibinfo {author} {\bibfnamefont {B.-B.}\ \bibnamefont {Ruan}},
  \bibinfo {author} {\bibfnamefont {J.}~\bibnamefont {Yu}}, \bibinfo {author}
  {\bibfnamefont {B.-J.}\ \bibnamefont {Pan}}, \bibinfo {author} {\bibfnamefont
  {Q.-G.}\ \bibnamefont {Mu}}, \bibinfo {author} {\bibfnamefont
  {T.}~\bibnamefont {Liu}}, \bibinfo {author} {\bibfnamefont {G.-F.}\
  \bibnamefont {Chen}},\ and\ \bibinfo {author} {\bibfnamefont {Z.-A.}\
  \bibnamefont {Ren}},\ }\bibfield  {title} {\bibinfo {title}
  {{Superconductivity in the ternary iridium–arsenide BaIr$_2$As$_2$}},\
  }\href {https://doi.org/10.1088/1361-6668/aa5298} {\bibfield  {journal}
  {\bibinfo  {journal} {Superconductor Science and Technology}\ }\textbf
  {\bibinfo {volume} {30}},\ \bibinfo {pages} {035007} (\bibinfo {year}
  {2017})}\BibitemShut {NoStop}%
\bibitem [{\citenamefont {Anand}\ \emph {et~al.}(2014)\citenamefont {Anand},
  \citenamefont {Kim}, \citenamefont {Tanatar}, \citenamefont {Prozorov},\ and\
  \citenamefont {Johnston}}]{2014ANAND_JOHNSTON}%
  \BibitemOpen
  \bibfield  {author} {\bibinfo {author} {\bibfnamefont {V.~K.}\ \bibnamefont
  {Anand}}, \bibinfo {author} {\bibfnamefont {H.}~\bibnamefont {Kim}}, \bibinfo
  {author} {\bibfnamefont {M.~A.}\ \bibnamefont {Tanatar}}, \bibinfo {author}
  {\bibfnamefont {R.}~\bibnamefont {Prozorov}},\ and\ \bibinfo {author}
  {\bibfnamefont {D.~C.}\ \bibnamefont {Johnston}},\ }\bibfield  {title}
  {\bibinfo {title} {{Superconductivity and physical properties of
  CaPd$_2$Ge$_2$ single crystals}},\ }\href
  {https://doi.org/10.1088/0953-8984/26/40/405702} {\bibfield  {journal}
  {\bibinfo  {journal} {Journal of Physics: Condensed Matter}\ }\textbf
  {\bibinfo {volume} {26}},\ \bibinfo {pages} {405702} (\bibinfo {year}
  {2014})}\BibitemShut {NoStop}%
\bibitem [{\citenamefont {Chajewski}\ \emph {et~al.}(2019)\citenamefont
  {Chajewski}, \citenamefont {Wi{\'{s}}niewski}, \citenamefont {Gnida},
  \citenamefont {Pikul},\ and\ \citenamefont {Kaczorowski}}]{2019GC_DK}%
  \BibitemOpen
  \bibfield  {author} {\bibinfo {author} {\bibfnamefont {G.}~\bibnamefont
  {Chajewski}}, \bibinfo {author} {\bibfnamefont {P.}~\bibnamefont
  {Wi{\'{s}}niewski}}, \bibinfo {author} {\bibfnamefont {D.}~\bibnamefont
  {Gnida}}, \bibinfo {author} {\bibfnamefont {A.~P.}\ \bibnamefont {Pikul}},\
  and\ \bibinfo {author} {\bibfnamefont {D.}~\bibnamefont {Kaczorowski}},\
  }\bibfield  {title} {\bibinfo {title} {{Crystal Growth and Physical
  Properties of the YPd$_2$Si$_2$ Superconductor}},\ }\href
  {https://doi.org/10.1021/acs.cgd.8b01386} {\bibfield  {journal} {\bibinfo
  {journal} {Crystal Growth and Design}\ }\textbf {\bibinfo {volume} {19}},\
  \bibinfo {pages} {2557} (\bibinfo {year} {2019})}\BibitemShut {NoStop}%
\bibitem [{\citenamefont {Han}\ \emph {et~al.}(2011)\citenamefont {Han},
  \citenamefont {Zhu}, \citenamefont {Mu}, \citenamefont {Zeng}, \citenamefont
  {Cheng}, \citenamefont {Shen},\ and\ \citenamefont {Wen}}]{2011FH_HHW}%
  \BibitemOpen
  \bibfield  {author} {\bibinfo {author} {\bibfnamefont {F.}~\bibnamefont
  {Han}}, \bibinfo {author} {\bibfnamefont {X.}~\bibnamefont {Zhu}}, \bibinfo
  {author} {\bibfnamefont {G.}~\bibnamefont {Mu}}, \bibinfo {author}
  {\bibfnamefont {B.}~\bibnamefont {Zeng}}, \bibinfo {author} {\bibfnamefont
  {P.}~\bibnamefont {Cheng}}, \bibinfo {author} {\bibfnamefont
  {B.}~\bibnamefont {Shen}},\ and\ \bibinfo {author} {\bibfnamefont {H.~H.}\
  \bibnamefont {Wen}},\ }\bibfield  {title} {\bibinfo {title} {{Absence of
  superconductivity in LiCu$_2$P$_2$}},\ }\href
  {https://doi.org/10.1021/ja108515f} {\bibfield  {journal} {\bibinfo
  {journal} {Journal of the American Chemical Society}\ }\textbf {\bibinfo
  {volume} {133}},\ \bibinfo {pages} {1751} (\bibinfo {year}
  {2011})}\BibitemShut {NoStop}%
\bibitem [{\citenamefont {Li}\ \emph {et~al.}(2018)\citenamefont {Li},
  \citenamefont {Ravichandran}, \citenamefont {Lindsay},\ and\ \citenamefont
  {Broido}}]{2018Li_Broido}%
  \BibitemOpen
  \bibfield  {author} {\bibinfo {author} {\bibfnamefont {C.}~\bibnamefont
  {Li}}, \bibinfo {author} {\bibfnamefont {N.~K.}\ \bibnamefont
  {Ravichandran}}, \bibinfo {author} {\bibfnamefont {L.}~\bibnamefont
  {Lindsay}},\ and\ \bibinfo {author} {\bibfnamefont {D.}~\bibnamefont
  {Broido}},\ }\bibfield  {title} {\bibinfo {title} {Fermi surface nesting and
  phonon frequency gap drive anomalous thermal transport},\ }\href
  {https://doi.org/10.1103/PhysRevLett.121.175901} {\bibfield  {journal}
  {\bibinfo  {journal} {Phys. Rev. Lett.}\ }\textbf {\bibinfo {volume} {121}},\
  \bibinfo {pages} {175901} (\bibinfo {year} {2018})}\BibitemShut {NoStop}%
\bibitem [{Note2()}]{Note2}%
  \BibitemOpen
  \bibinfo {note} {The only difference between CaPd$_2$Si$_2$ and CaPd$_2$P$_2$
  lies in the Si/P site. The atomic number of Si is 14, whereas that of P is
  15.}\BibitemShut {Stop}%
\bibitem [{\citenamefont {Pauling}(1960)}]{1960PAULING}%
  \BibitemOpen
  \bibfield  {author} {\bibinfo {author} {\bibfnamefont {L.}~\bibnamefont
  {Pauling}},\ }\href@noop {} {\emph {\bibinfo {title} {The Nature of the
  Chemical Bond}}},\ \bibinfo {edition} {3rd}\ ed.\ (\bibinfo  {publisher}
  {Cornell University Press},\ \bibinfo {address} {Ithaca, NY},\ \bibinfo
  {year} {1960})\ p.~\bibinfo {pages} {93}\BibitemShut {NoStop}%
\bibitem [{\citenamefont {Hadi}\ \emph {et~al.}(2017)\citenamefont {Hadi},
  \citenamefont {Ali}, \citenamefont {Naqib},\ and\ \citenamefont
  {Islam}}]{2017Hadi_Islam}%
  \BibitemOpen
  \bibfield  {author} {\bibinfo {author} {\bibfnamefont {M.~A.}\ \bibnamefont
  {Hadi}}, \bibinfo {author} {\bibfnamefont {M.~S.}\ \bibnamefont {Ali}},
  \bibinfo {author} {\bibfnamefont {S.~H.}\ \bibnamefont {Naqib}},\ and\
  \bibinfo {author} {\bibfnamefont {A.~K. M.~A.}\ \bibnamefont {Islam}},\
  }\bibfield  {title} {\bibinfo {title} {New ternary superconducting compound
  laru2as2: Physical properties from density functional theory calculations},\
  }\href {https://doi.org/10.1088/1674-1056/26/3/037103} {\bibfield  {journal}
  {\bibinfo  {journal} {Chinese Physics B}\ }\textbf {\bibinfo {volume} {26}},\
  \bibinfo {pages} {037103} (\bibinfo {year} {2017})}\BibitemShut {NoStop}%
\end{thebibliography}%
\end{document}